\documentclass[a4paper,showpacs,preprintnumbers,amsmath,amssymb,floatfix]{revtex4}
\usepackage[german, english]{babel}
\usepackage{amsmath}
\usepackage{amssymb}
\usepackage{graphicx}
\usepackage{epsfig}
\usepackage{color}
\usepackage[hang,nooneline]{subfigure}
\newcounter{fig}

\newcommand{\hq}{q}
\newcommand{\cs}{\cos\sigma}

\begin{document}

\title{Spinning Wormholes in Scalar-Tensor Theory}
%\vspace{1.5truecm}
\author{Xiao Yan Chew}
\email[{\it Email:}]{xiao.yan.chew@uni-oldenburg.de}
\author{Burkhard Kleihaus}
\email[{\it Email:}]{b.kleihaus@uni-oldenburg.de}
\author{Jutta Kunz}
\email[{\it Email:}]{jutta.kunz@uni-oldenburg.de}
\affiliation{
Institut f\"ur  Physik, Universit\"at Oldenburg, Postfach 2503,
  D-26111 Oldenburg, Germany}

\date{\today}

\begin{abstract}
We consider spinning generalizations of the Ellis wormhole in 
scalar-tensor theory. Analogous to other compact objects
%neutron stars or boson stars,
these wormholes can carry a non-trivial scalarization.
We determine the domain of existence of the scalarized wormholes
and investigate the effect of the scalarization on their
properties.
Depending on the choice of the coupling function, they may
possess multiple throats and equators in the Jordan frame,
while possessing only a single throat in the Einstein frame.
\end{abstract}

\maketitle

\section{Introduction}

Among the contenders of General Relativity (GR)
scalar-tensor theories (STT) 
\cite{Jordan:1949zz,Fierz:1956zz,Jordan:1959eg,Brans:1961sx,Dicke:1961gz,Bergmann:1968ve,Wagoner:1970vr}
hold a prominent place
(see e.g.~\cite{Fujii:2003pa,Faraoni:2004pi} for reviews).
When considering besides the gravitational tensor field 
the presence of an additional gravitational scalar field,
the formulations of STT are
usually restricted by a number of physical requirements.
In particular, the STT should obey the well-known observational
constraints.

STT predict a number of new phenomena, not present in GR.
One such phenomenon is gravitational dipole radiation, which would be 
emitted, for instance, from inspiralling close binaries 
\cite{Eardley:1975,Will:1989sk}.
Another such phenomenon is the existence of compact solutions,
which possess a finite gravitational scalar field, when the
coupling exceeds a critical strength.
Dubbed {\sl spontaneous scalarization}, this effect was first
observed for static neutron stars \cite{Damour:1993hw,Damour:1996ke}
(see also
\cite{Harada:1998ge,Harada:1997mr,Salgado:1998sg,Sotani:2012eb,Pani:2014jra,Silva:2014fca,Sotani:2017pfj,Motahar:2017blm})
and recently demonstrated for rapidly rotating neutron stars
\cite{Doneva:2013qva,Doneva:2014uma,Doneva:2014faa,Staykov:2016mbt}.
Spontaneous scalarization is also known to occur in boson stars 
\cite{Whinnett:1999sc,Alcubierre:2010ea,Ruiz:2012jt,Kleihaus:2015iea}
and hairy black holes
\cite{Kleihaus:2015iea}.
Here we show, that scalarization also arises for wormholes.

Wormholes represent intriguing topologically non-trivial solutions, 
connecting either two asymptotically flat universes by a throat
or connecting two distant regions within a single universe.
Whereas the non-traversable 
Einstein-Rosen bridge \cite{Einstein:1935tc} of GR
represents a feature of the Schwarz\-schild spacetime,
traversable wormholes in GR need exotic matter for their existence
\cite{Morris:1988cz,Visser}.
The simplest such traversable wormholes 
based on phantom fields are the static Ellis wormholes
\cite{Ellis:1973yv,Ellis:1979bh,Bronnikov:1973fh}.
We note that exotic matter in the form of
phantom fields can be employed in cosmology
to model the accelerated expansion of the Universe
(see e.g.~\cite{Lobo:2005us}).

Rotating generalizations of the Ellis wormholes 
have only been found recently.
These include analytically constructed
slowly rotating perturbative wormhole solutions
\cite{Kashargin:2007mm,Kashargin:2008pk}
as well as rapidly rotating non-perturbative solutions
\cite{Kleihaus:2014dla,Chew:2016epf},
that were obtained numerically.
The rotating wormhole metric presented by Teo \cite{Teo:1998dp},
however, does not represent a solution of a specified set of
Einstein-matter equations.

While wormholes represent hypothetical objects,
they have been searched for observationally
\cite{Abe:2010ap,Toki:2011zu,Takahashi:2013jqa},
and a number of their observational signatures have been 
addressed already, such as their gravitational lensing effects
\cite{Cramer:1994qj,Safonova:2001vz,Perlick:2003vg,Nandi:2006ds,Nakajima:2012pu,Kuhfittig:2013hva,Tsukamoto:2016zdu,Shaikh:2017zfl} 
including their Einstein rings \cite{Tsukamoto:2012xs},
their shadows \cite{Bambi:2013nla,Nedkova:2013msa},
or their accretion disks \cite{Zhou:2016koy}.
Also combined neutron star--wormhole systems
(see e.g.~\cite{Aringazin:2014rva,Dzhunushaliev:2016ylj}
and references therein)
and boson star--wormhole systems 
\cite{Dzhunushaliev:2014bya,Dzhunushaliev:2017syc,Hoffmann:2017jfs,Hoffmann:2017vkf}
have been considered.

Here we construct rotating wormhole solutions in STT,
which are based on the presence of a phantom field.
Thus they fundamentally differ from wormholes obtained previously in STT,
which were pure (static) STT solutions without any further (exotic) matter 
fields present
\cite{Agnese:1995kd,Nandi:1997mx,Nandi:1997en,Eiroa:2008hv,Bronnikov:2009tv,Bhattacharya:2009rt,Lobo:2010sb,Bronnikov:2010tt,Sushkov:2011zh,Yue:2011cq,Shaikh:2016dpl}.
Our main insight here consists in the realization
that once the wormhole solutions are known in GR,
analytically or numerically, then in order to obtain
scalarized wormhole solutions in the Einstein frame,
only  the two scalar field equations must be considered.
In fact, a single solution for a wormhole metric in GR
gives rise to a whole family of STT wormhole solutions
in the Einstein frame,
which differ only in their scalar fields.

In the transformation between the Einstein frame and the Jordan
frame the nonminimal coupling ${\cal A}$ plays a major role.
It represents the interaction of the gravitational scalar field
with the matter fields, i.e., here with the phantom  field.
Clearly  the wormhole solutions should possess a strong dependence
on the choice of this coupling. To illustrate this dependence,
we present wormhole solutions for 3 examples of the coupling ${\cal A}$.
The first example corresponds to the one employed in the
discovery of the scalarization of neutron  stars \cite{Damour:1993hw,Damour:1996ke},
where we obtain particularly interesting scalarized wormhole solutions
for positive values of the coupling constant (when there is no
scalarization in neutron stars).
These wormholes may possess many equators and throats in the Jordan frame.
The second example corresponds to Brans-Dicke theory
\cite{Brans:1961sx},
while the third example has been inspired by \cite{Barcelo:1999hq}.
While in the first and third example 
the rotating wormhole solutions of GR \cite{Kleihaus:2014dla,Chew:2016epf}
are also solutions of the STT equations, this is not the case
in the second example.

%But the presence of the gravitational scalar field together with
%the phantom field gives rise to wormhole solutions with
%scalarization. 
%We show that in constrast to neutron stars or boson stars,
%there is no lower limit on the magnitude of the scalar coupling
%constant for scalarization to occur.

In section II we state the theoretical setting.
We briefly recall the action for scalar-tensor theories,
and discuss the transition from the Jordan to the Einstein frame.
We present the ansatz, the boundary
conditions and the equations of motion.
We then determine the domain of existence for the scalar charges
and address the mass, the angular momentum, and the
quadrupole moment as well as
the geometric properties and the null energy condition.
The results for the scalarized wormholes
for the 3 examples of the nonminimal coupling ${\cal A}$ are presented in section III,
while section IV gives our conclusions.

\section{Theoretical Setting}

\subsection{Scalar-Tensor Theories}

Let us consider STT with a single gravitational scalar field $\Phi$.
The most general action giving rise to second order field equations
then contains three functions of the gravitational scalar field $\Phi$,
$F(\Phi)$, $Z(\Phi)$, and $W(\Phi)$,
and reads in the (physical) Jordan frame
\begin{eqnarray} \label{JFA}
S = \frac{1}{16\pi G_{*}} \int d^4x \sqrt{-{\tilde g}}
\left({F(\Phi)\tilde R} 
- Z(\Phi){\tilde g}^{\mu\nu}\partial_{\mu}\Phi\partial_{\nu}\Phi   
-2 W(\Phi) \right) +
S_{m}\left[\Psi_{m};{\tilde g}_{\mu\nu}\right] .
\end{eqnarray}
Here 
$G_{*}$ is the bare gravitational constant,
${\tilde g}_{\mu\nu}$ is the spacetime metric, 
and ${\tilde R}$ is the Ricci scalar curvature. 
The matter fields $\Psi_{m}$ are contained in the 
$S_{m}\left[\Psi_{m};{\tilde g}_{\mu\nu}\right]$ part of the action,
which depends on the space-time metric ${\tilde g}_{\mu\nu}$.
It does not involve the gravitational scalar field $\Phi$
to satisfy the weak equivalence principle.

The functions $F(\Phi)$ and $Z(\Phi)$ 
should satisfy a set of physical restrictions.
We require that $F(\Phi)>0$, 
since gravitons should carry positive energy,
while the requirement
$2F(\Phi)Z(\Phi) + 3[dF(\Phi)/d\Phi]^2 \ge 0$ guarantees, that
the kinetic energy of the gravitational scalar field is not negative.
When the potential function $W(\Phi)$ is chosen to vanish,
this would correspond to a massless scalar field without self-interaction.

The gravitational and matter field equations in the Jordan frame
are obtained by varying the action with respect to the metric components,
the gravitational scalar field  and the matter fields.
This procedure then leads to a rather intricate set of
coupled field equations.
Alternatively, one may consider a mathematically equivalent
approach to STT, obtained by invoking
the conformally related Einstein frame with the metric $g_{\mu\nu}$
\begin{equation}\label {CONF1}
g_{\mu\nu} = F(\Phi){\tilde g}_{\mu\nu} .
\end{equation}
Substitution of $g_{\mu\nu}$ in the action Eq.~(\ref{JFA}) 
leads to the action in
the Einstein frame, which reads (up to a boundary term)
\begin{eqnarray}
S= \frac{1}{16\pi G_{*}}\int d^4x \sqrt{-g} \left(R -
2g^{\mu\nu}\partial_{\mu}\varphi \partial_{\nu}\varphi -
4V(\varphi)\right)+ S_{m}[\Psi_{m}; {\cal A}^{2}(\varphi)g_{\mu\nu}] .
\end{eqnarray}
Here $R$ is the Ricci scalar curvature with respect to the Einstein
metric $g_{\mu\nu}$.
The new gravitational scalar field $\varphi$ is defined via
\begin{equation}\label {CONF2}
\left(d\varphi \over d\Phi \right)^2 = {3\over
4}\left({d\ln(F(\Phi))\over d\Phi } \right)^2 + {Z(\Phi)\over 2
F(\Phi)}
\end{equation}
and the new functions ${\cal A}(\varphi)$ and $V(\varphi)$ are given by
\begin{equation}\label{CONF3}
{\cal A}(\varphi) = F^{-1/2}(\Phi) \, , \qquad
2V(\varphi) = W(\Phi)F^{-2}(\Phi).
\end{equation}
As a consequence of the transformation,
the gravitational scalar field appears in the matter action 
in the Einstein frame via the nonmiminal coupling
${\cal A}^{2}(\varphi)$.

\subsection{Action and Ansatz}

We now turn to the construction of wormhole solutions 
in STT, supported by a phantom field $\Psi$ as the matter field. 
We further assume the potential $V(\varphi)$ to vanish.
Thus both scalar fields have only a kinetic term in the action,
while they interact via the nonminimal coupling ${\cal A}(\varphi)$.
In the Einstein frame the action then reads
\begin{eqnarray}
S= {1\over 16\pi G_{*}}\int d^4x \sqrt{-g} \left(R -
2g^{\mu\nu}\partial_{\mu}\varphi \partial_{\nu}\varphi 
  +8\pi G_{*} {\cal A}^{2} g^{\mu\nu}
  \partial_{\mu}\Psi \partial_{\nu}\Psi
  \right)
\end{eqnarray}

In the following we change to the scaled phantom field 
$\psi =\sqrt{4\pi G_{*}}\Psi $.
The Einstein equations can then be cast in the form
\begin{equation}
R_{\mu\nu} = 2\left( \partial_\mu \varphi \partial_\nu \varphi 
         -{\cal A}^2\partial_\mu \psi\partial_\nu \psi
	 \right) \ ,
\label{einsteinequ}	 
\end{equation}
and the field equations for the scalar fields yield
\begin{eqnarray}
\partial_\mu\left(\sqrt{-g}g^{\mu\nu}{\cal A}^2\partial_\nu \psi\right) 
& = & 0 \ , 
\label{equpsi}
\\
\partial_\mu\left(\sqrt{-g}g^{\mu\nu}\partial_\nu \varphi\right) 
& = & -\frac{1}{2}\frac{d {\cal A}^2}{d\varphi}\sqrt{-g} \partial_\lambda\psi\partial^\lambda\psi \  .
\label{equphi}
\end{eqnarray}

For stationary rotating spacetimes we employ the line element
\begin{equation}
ds^2 = -e^{f} dt^2 +p^2 e^{-f} 
\left( e^{\nu} \left[d\eta^2 +\hq d\theta^2\right]
                    + \hq \sin^2\theta \left(d\phi -\omega dt\right)^2\right) \ .
\label{lineel}
\end{equation}
Here the functions $f$, $p$, $\nu$ and $\omega$ depend only on the coordinates
$\eta$ and $\theta$, and $\hq=\eta^2 + \eta_0^2$ 
is an auxiliary function. 
The radial coordinate $\eta$ takes positive and negative
values,  covering the real line $-\infty< \eta < \infty$. 
In the limits $\eta\to \pm\infty$
two distinct asymptotically flat regions $\Sigma_\pm$ are approached.
The two scalar fields $\varphi$ and $\psi$ depend only on the coordinates
$\eta$ and $\theta$, as well.

\subsection{Equations of Motion and Boundary Conditions}

By substituting the Ansatz (\ref{lineel}) 
into the general set of equations of motion
a system of non-linear partial differential equations (PDEs) is obtained.
As noted before \cite{Kleihaus:2014dla},
the PDE for the metric function $p$ decouples
and is given by
\begin{equation}
\partial_\eta^2 p +\frac{3\eta}{\hq}\partial_\eta p + 
\frac{2 \cos\theta}{\hq\sin\theta}\partial_\theta p+
\frac{1}{\hq}\partial_\theta^2 p = 0 \ .
\label{pde_p}
\end{equation}
Imposing the boundary conditions 
$p(\eta \to \infty) =p(\eta \to -\infty) = 1$ and 
$\partial_\theta p(\theta =0) =\partial_\theta p(\theta =\pi) = 0$,
a trivial solution of this equation is given by %$p=1$.
\begin{equation}
p=1 \ .
\label{p=1}
\end{equation}

Inserting the resulting metric Ansatz into the phantom field equation
(\ref{equpsi}) leads to
\begin{equation}
\partial_\eta\left( \hq \sin\theta {\cal A}^2 \partial_\eta \psi\right) 
+\partial_\theta\left(\sin\theta {\cal A}^2 \partial_\theta \psi\right) 
=0 \ .
\label{pde_psi}
\end{equation}
Assuming
$\partial_\theta \psi=0$, 
a first integral is obtained
\begin{equation}
\partial_\eta \psi = \frac{Q_\psi {\cal A}^{-2}}{\hq}\ .
\label{sol_psie}
\end{equation}
We will assume that in the asymptotic region $\Sigma_+$
the gravitational scalar field vanishes, $\varphi(+\infty)=0$, and ${\cal A}(0)=1$. Then the integration
constant $Q_\psi$ can be identified with the phantom scalar charge.
We will also assume that the phantom field vanishes in the region $\Sigma_+$,
i.~e.~$\psi(+\infty)=0$.

Addressing next the gravitational scalar field $\varphi$, 
we now assume that $\partial_\theta \varphi=0$, as well. Thus Eq.~(\ref{equphi})
reduces to 
\begin{equation}
\partial_\eta \left(\hq \partial_\eta \varphi\right) = 
-\frac{1}{2}\frac{1}{{\cal A}^4}\frac{d {\cal A}^2}{d\varphi} \frac{1}{\hq}Q_\psi^2 \ .
\label{redequphi}
\end{equation}

Let us next turn to the Einstein equations.
Inserting $p=1$, $\partial_\theta \psi=0$ and $\partial_\theta \varphi=0$
shows that the Einstein equations $R_{\phi\phi}=0$, $R_{\theta\theta}=0$,
$R_{t\phi}=0$, $R_{tt}=0$ and $R_{\eta\theta}=0$
are independent of the scalar fields.
These equations lead to three second order PDEs 
for the metric functions $f$, $\omega$ and $\nu$, and to a constraint,
\begin{eqnarray}
0 & = & 
 \partial_\eta\left( \hq \sin\theta \partial_\eta f\right)
+\partial_\theta\left( \sin\theta \partial_\theta f\right)
-\hq \sin^3\theta e^{-2 f}\left(\hq (\partial_\eta\omega)^2 
                              + (\partial_\theta\omega)^2 \right)
\label{pdef}\\
0 & = & 
 \partial_\eta\left(\hq^2\sin^3\theta e^{-2 f} \partial_\eta\omega\right)		      
+\partial_\theta\left(\hq\sin^3\theta e^{-2 f} \partial_\theta\omega\right)		      
\label{pdeo}\\
0 & = & 
 \partial_\eta\left( \hq \sin\theta \partial_\eta \nu\right)
 +\sin\theta \partial_{\theta\theta} \nu -\cos\theta \partial_\theta\nu
 -\hq \sin^3\theta e^{-2 f}\left(\hq (\partial_\eta\omega)^2 
                              -2 (\partial_\theta\omega)^2 \right)
\label{pdenu}\\
0 & = & 
-\hq \sin\theta \partial_\eta f \partial_\theta f
+ \hq \cos\theta \partial_\eta \nu + \eta\sin\theta \partial_\theta \nu
+ \hq^2\sin^3\theta e^{-2 f}\partial_\eta \omega \partial_\theta \omega
\label{constraint}
\end{eqnarray}

Let us now consider the boundary conditions,
which should be imposed
in the asymptotic regions $\Sigma_\pm$ and on the axis 
$\theta = 0, \pi$.
In the  asymptotic region $\Sigma_+$, i.e., 
for $\eta \to +\infty$,
we require that the metric approaches the Minkowski spacetime
\begin{equation}
\left. f\right|_{\eta \to \infty} = 0 \ , \ 
\left. \omega\right|_{\eta \to \infty} = 0 \ , \ 
\left. \nu\right|_{\eta \to \infty} = 0 \ .
\label{bcinfty}
\end{equation}
In  asymptotic region $\Sigma_-$, i.e.,
for $\eta \to -\infty$ we allow for finite values of the
functions $f$ and $\omega$,
\begin{equation}
\left. f\right|_{\eta \to -\infty} = \gamma \ , \ 
\left. \omega\right|_{\eta \to -\infty} = \omega_{-\infty} \ , \ 
\left. \nu\right|_{\eta \to -\infty} = 0 \ .
\label{bcmininfty}
\end{equation}
%The condition on $\nu$ ensures that the spacetime has no conical singularity.

The parameter $\gamma$ controls the symmetry of the wormhole solutions.
We call the solutions symmetric, when $\gamma=0$,
and non-symmetric, when $\gamma \ne 0$.
Therefore we refer to $\gamma$ as the asymmetry parameter.
The parameter $\omega_{-\infty}$ controls the rotation of the spacetime.
For static wormhole solutions $\omega_{-\infty}=0$. 
Static wormholes are known in closed form,
\begin{equation}
f = \frac{\gamma}{2} \left(1-\frac{2}{\pi}\arctan\left(\frac{\eta}{\eta_0}\right) \right) \ , \
\omega = 0  \ , \ \nu = 0  \ . \
\label{statsol}
\end{equation}
We note, that in order to obtain the Minkowski spacetime
in the limit $\eta \to -\infty$, 
a suitable coordinate transformation needs to be performed
(see below in subsection \ref{sigma_minus}).

The last set of boundary conditions concerns the symmetry axis.
Here regularity requires
\begin{equation}
\left.\partial_\theta f\right|_{\theta = 0} = 0 \ , \ 
\left.\partial_\theta \omega \right|_{\theta = 0} = 0 \ , \ 
\left. \nu\right|_{\theta = 0} = 0 \ ,
\label{bcaxis}
\end{equation}
together with the analogous conditions for $\theta = \pi$.

%Rotating wormhole solutions in GR
%have been studied in \cite{Kleihaus:2014dla,Chew:2016epf}.
We have not yet addressed
the remaining Einstein equation for $R_{\eta\eta}$.
Substituting the solutions for $f$, $\omega$ and $\nu$ in $R_{\eta\eta}$ shows,
that it satisfies
\begin{equation}
R_{\eta\eta} = - 2\frac{D^2}{\hq^2} \ ,
\label{sol_psie2}
\end{equation}
where the constant $D$ depends on the mass and the angular momentum
of the spacetime.
Explicitly we find
\begin{equation}
D^2  =  \frac{\hq}{4}\left[ \hq  (\partial_\eta f)^2 - (\partial_\theta f)^2 \right]
         -\frac{\hq}{2}\left(\eta \partial_\eta \nu-
	                  \frac{\cos\theta}{\sin\theta}\partial_\theta \nu\right)
         -\frac{\hq^2}{4}\sin^2\theta e^{-2 f}
	 \left[ \hq (\partial_\eta\omega)^2 
               - (\partial_\theta\omega)^2 \right] +\eta_0^2 \ .
\label{eqd2}
\end{equation}
In the pure Einstein case and for vanishing gravitational scalar field $\varphi$,
the constant $D$ is simply given by the phantom scalar charge $Q_\psi$, as seen by
substituting Eq.~(\ref{sol_psie}) into the right hand side (rhs) of the Einstein equation.
Since the scalar fields do not enter the PDSs for $f$, $\omega$ and $\nu$,
the rhs of Eq.~(\ref{sol_psie2}) must retain this constant
also in the presence of both scalar fields.
%
%which takes values from $D=\eta_0$ for the 
%static solutions and $D=0$ in the extremal rotating limit.
Consequently,
\begin{equation}
- \frac{D^2}{\hq^2} =  \partial_\eta \varphi \partial_\eta \varphi 
                            -{\cal A}^{-2}\frac{Q_\psi^2}{\hq^2} \ .
\label{redeqRee2}
\end{equation}
This leads to the first order ODE for the gravitational scalar field $\varphi$
\begin{equation}
\partial_\eta \varphi= \pm \frac{1}{\hq}\sqrt{{\cal A}^{-2}Q_\psi^2-D^2} \ .
\label{equphi1}
\end{equation}
We note that any solution of Eq.~(\ref{equphi1}) is also a solution of
the second order equation (\ref{redequphi}).

From Eq.~(\ref{equphi1}) we can read off the scalar charge of the gravitional scalar field,
$Q_\varphi = \pm\sqrt{Q_\psi^2-D^2}$.
Hence, for any wormhole spacetime the two scalar charges are related by
\begin{equation}
Q_\psi^2-Q_\varphi^2=D^2 \ ,
\label{Qrelation}
\end{equation}
where the quantity $D$ depends on the mass and angular momentum of the 
spacetime.

\subsection{Mass and  Angular Momentum}

Let us next address the mass and the angular momentum 
of the wormholes, which should be obtained in the physical Jordan frame.
In general, STT give rise to different types of mass, such as
the gravitational mass $M_{\rm K}$, the tensor mass $M_{\rm T}$ or the Schwarzschild mass
$M_{\rm S}$
(see e.g.~\cite{Lee:1974pt,Scheel:1994yr,Scheel:1994yn,Whinnett:1999ws,Yazadjiev:1999hy}).
The tensor mass simply corresponds to the ADM mass in the Einstein frame,
$M_{\rm T}=M_{\rm E}$.
It has appealing properties, such as being positive definite,
or exhibiting a monotonic decrease in the emission of gravitational waves
\cite{Lee:1974pt,Scheel:1994yr,Scheel:1994yn}.
In the following we consider the mass and angular momentum first
in the asymptotic region $\Sigma_+$ and then in $\Sigma_-$.

\subsubsection{Asymptotic region $\Sigma_+$}

Let us first recall the mass $M_{E+}$ and the angular momentum $J_+$ in the Einstein frame.
They are encoded in the asymptotic behavior of the metric functions
$f(\eta)$ and $\omega(\eta)$ in the asymptotic region $\Sigma_+$.
Therefore we can read off $M_{E+}$ and $J_+$ in the Einstein frame directly,
\begin{eqnarray}
& &   
f \underset{\eta \to + \infty} 
\longrightarrow -\frac{2 M_{E+}}{\eta} \ , \ \ \ 
\omega \underset{\eta \to + \infty} 
\longrightarrow \frac{2J_+}{\eta^3} \ . 
\label{asymp1} 
\end{eqnarray}

Let us now turn to the Jordan frame and find the relations between
the mass and the angular momentum in the two frames.
In order to express the masses $M_{\rm K}$ and $M_{\rm S}$ in the Jordan frame by the mass and
the scalar charge in the Einstein frame we consider the asymptotic behaviour
of the metric in the Jordan frame as $\eta \to \infty$,
\begin{eqnarray}
-\tilde{g}_{tt} & = & 1 -\frac{2 M_{\rm K+}}{\eta} + {\cal O}(\eta^{-2}) 
= {\cal A}(\varphi)^2\left(1 -\frac{2 M_{\rm E+}}{\eta}\right)+ {\cal O}(\eta^{-2})
\nonumber\\
& =& 
1 -\frac{2}{\eta}  
\left(M_{\rm E+}+\left.\frac{d{\cal A}}{d\varphi}\right|_{\varphi=0} Q_\varphi\right)
+ {\cal O}(\eta^{-2}) \ ,
\label{eqMk}\\
\tilde{g}_{\eta\eta} & = & 1 +\frac{2 M_{\rm S+}}{\eta} + {\cal O}(\eta^{-2}) 
= {\cal A}(\varphi)^2\left(1 +\frac{2 M_{\rm E+}}{\eta}\right)+ {\cal O}(\eta^{-2})
\nonumber\\
& = & 
1 +\frac{2}{\eta} 
\left(M_{\rm E+}-\left.\frac{d{\cal A}}{d\varphi}\right|_{\varphi=0} Q_\varphi\right)
+ {\cal O}(\eta^{-2}) \ ,
\label{eqMs}
\end{eqnarray}
where we used the expansions 
\begin{equation}
{\cal A}(\varphi)^2 = 
1 + 2 \left.\frac{d{\cal A}}{d\varphi}\right|_{\varphi=0}\varphi 
+ {\cal O}(\varphi^2)
=1 - 2 \left.\frac{d{\cal A}}{d\varphi}\right|_{\varphi=0}\frac{Q_\varphi}{\eta}
+ {\cal O}(\eta^{-2})
%\ , \ \ \ {\rm with} \ \ \varphi= -\frac{Q_\varphi}{\eta}+ {\cal O}(\eta^{-2}) \ . 
\label{exFinfty}
\end{equation}
with $\varphi= -Q_\varphi/\eta+ {\cal O}(\eta^{-2})$.
Hence we find  for the gravitational mass $M_{\rm K+}$ and the Schwarzschild mass $M_{\rm S+}$
\begin{eqnarray}
M_{\rm K+} & = & M_{\rm E+}+ \left.\frac{d{\rm ln}{\cal A}}{d\varphi}\right|_{\varphi=0} Q_\varphi\ ,
\label{MkrelMe}\\
M_{\rm S+} & = & M_{\rm E+}- \left.\frac{d{\rm ln}{\cal A}}{d\varphi}\right|_{\varphi=0} Q_\varphi \ ,
\label{MsrelMe}
\end{eqnarray}
respectively, where we used 
$\left.\frac{d{\cal A}}{d\varphi}\right|_{\varphi=0}=
\left.\frac{d{\rm ln}{\cal A}}{d\varphi}\right|_{\varphi=0}$ (since  ${\cal A}(0)=1$) for
convenience.
We read off the simple relations 
\begin{equation}
M_{\rm K+}+ M_{\rm S+} = 2 M_{\rm T+} \ , \ \ \ 
M_{\rm K+}- M_{\rm S+} = 2\left.\frac{d{\rm ln}{\cal A}}{d\varphi}\right|_{\varphi=0} Q_\varphi \ .
\label{Jmassrel}
\end{equation}
We observe that the gravitational mass, the Schwarzschild mass and the tensor mass
coincide in the Jordan frame (and also coincide with the ADM mass in the Einstein frame),
provided $\left.\frac{d{\cal A}}{d\varphi}\right|_{\varphi=0}=0$.

%For the STT considered here, where ${\cal A}$ is given by eq.~(\ref{calA})
%and where the gravitational scalar field goes to zero asymptotically
%for $\eta \to 0$, the mass $M_{\rm E +}$ as read off at
%plus infinity in the Einstein frame agrees with the mass $M_{\rm J +}$
%in the Jordan frame \cite{Doneva:2013qva},
%\begin{equation}
%M_{\rm E +} = M_{\rm J +} = M_+ \ , \ \ \
%g_{tt} \underset{\eta \to + \infty} 
%\longrightarrow - \left(1 - \frac{2 M_+}{\eta}\right) 
%\label{mass}
%\end{equation}
The angular momentum, as read off at
plus infinity in the Einstein frame agrees with the angular momentum
in the Jordan frame \cite{Doneva:2013qva},
\begin{equation}
J_{\rm E +} = J_{\rm J +} = J_+ \ , \ \ \
\frac{g_{t\phi}}{g_{\phi\phi}} 
= \frac{\tilde{g}_{t\phi}}{\tilde{g}_{\phi\phi}} \underset{\eta \to +\infty} 
\longrightarrow -\frac{2 J_+}{\eta^3} \ .
\label{angmom}
\end{equation}

\subsubsection{Asymptotic region $\Sigma_-$}
\label{sigma_minus}

In the next step we consider the mass and the angular momentum
in the asymptotic region $\Sigma_-$, i.~e.~ $\eta \to -\infty$.
In this region the expansion in the Einstein frame reads
\begin{eqnarray}
& &
f \underset{\eta \to - \infty} 
\longrightarrow \gamma+\frac{2 M_{\rm E -}}{\eta}\ , \ \ \ 
\omega \underset{\eta \to - \infty} 
\longrightarrow \omega_{-\infty}+ \frac{2 J_{\rm E -}}{\eta^3} .
\label{asymp2}
\end{eqnarray}
Here, 
to identify the mass and the angular momentum $\bar M_{\rm E -}$ 
and $\bar J_{\rm E -}$ in the Einstein frame,
a coordinate transformation has to be performed
to obtain an asymptotically flat spacetime in this limit.
This is achieved by the transformation
\begin{eqnarray}
\label{trans}
\bar t = e^{\gamma/2} t \ , \ \ \
\bar \eta = e^{-\gamma/2} \eta \ , \ \ \
\bar \phi = \phi - \omega_{-\infty} t \ ,
%\nonumber\\
\end{eqnarray}
leading to
$\bar M_{\rm E -}$ and $\bar J_{\rm E -}$ in terms of the quantities
$M_{\rm E -}$ and $J_{\rm E -}$,
\begin{eqnarray}
\label{mom-}
\bar J_{\rm E -} = J_{\rm E -} e^{-2\gamma} \ , \ \ \
\bar M_{\rm E -} = M_{\rm E -} e^{-\gamma/2} \ .
\end{eqnarray}
In order to relate the mass and angular momentum in the
Jordan frame to the corresponding quantities in the Einstein frame 
we have to take into account that the gravitational scalar field
assumes a finite value for $\eta \to - \infty$. 
Let us define the asymptotic quantities 
$\varphi_- = \varphi(\eta \to -\infty)$ and ${\cal A}_{-}={\cal A}(\varphi_-)$.
The coordinate transformation which yields an asymptotically flat metric
is now given by
\begin{eqnarray}
\label{transJ}
\bar t = e^{\gamma/2} {\cal A}_{-} t \ , \ \ \
\bar \eta = e^{-\gamma/2} {\cal A}_{-} \eta \ , \ \ \
\bar \phi = \phi - \omega_{-\infty} t \ .
%\nonumber\\
\end{eqnarray}
In these coordinates
\begin{eqnarray}
\tilde{g}_{\bar t\bar t} =
\frac{{\cal A}^2}{{\cal A}_{-}^2} e^{f-\gamma}
\underset{\eta \to - \infty} \longrightarrow 
& &
\left(1+2\frac{1}{{\cal A}_{-}}\frac{d{\cal A}}{d\varphi}(\varphi-\varphi_-)\right)
\left(1+\frac{2 M_{\rm E-}}{\eta}\right) +{\cal O}(\eta^{-2})
\label{expgttJm}\\
& = &
\left(1+2\frac{d{\cal A}}{d\varphi}\frac{Q_{\varphi-}e^{-\gamma/2}}{\bar\eta}  \right)
\left(1+\frac{2 M_{\rm E-}e^{-\gamma/2} {\cal A}_{-}}{\bar\eta}\right) +{\cal O}(\bar\eta^{-2})
\label{expgttJma}\\
& = &
1 +\frac{2}{\bar\eta}
{\cal A}_{-}\left( M_{\rm E-}e^{-\gamma/2} +
\left[\frac{d{\rm ln}{\cal A}}{d\varphi}\right]_{\varphi_-}Q_{\varphi-}e^{-\gamma/2}
\right)
+{\cal O}(\bar\eta^{-2}) \ ,
\label{expgttJmb}
\end{eqnarray}
leading to
\begin{equation}
M_{\rm K-} ={\cal A}_{-}\left( M_{\rm E-}e^{-\gamma/2} +
\left[\frac{d{\rm ln}{\cal A}}{d\varphi}\right]_{\varphi_-}Q_{\varphi-}e^{-\gamma/2}
\right)
= {\cal A}_{-}\left(\bar{M}_{\rm E-}
 +
\left[\frac{d{\rm ln}{\cal A}}{d\varphi}\right]_{\varphi_-}\bar{Q}_{\varphi-}
\right) \ ,
\label{MKm}
\end{equation}
where we defined 
$\bar{Q}_{\varphi-} = e^{-\gamma/2}Q_{\varphi-}$ with
$Q_{\varphi -} =-(\eta^2\partial_{\eta}\varphi)_{-\infty}$.
Similarly we find 
\begin{equation}
M_{\rm S-} = {\cal A}_{-}\left( M_{\rm E-}e^{-\gamma/2} -
\left[\frac{d{\rm ln}{\cal A}}{d\varphi}\right]_{\varphi_-}Q_{\varphi-}e^{-\gamma/2}
\right)
= {\cal A}_{-}\left(\bar{M}_{\rm E-}
 -
\left[\frac{d{\rm ln}{\cal A}}{d\varphi}\right]_{\varphi_-}\bar{Q}_{\varphi-}
\right)\ .
\label{MKs}
\end{equation}
We read off the simple relations 
\begin{equation}
M_{\rm K-}+ M_{\rm S-} = 2 {\cal A}_{-}\bar{M}_{\rm E-} \ , \ \ \ 
M_{\rm K-}- M_{\rm S-} = 
2{\cal A}_{-}\left.\frac{d{\rm ln}{\cal A}}{d\varphi}\right|_{\varphi_-} \bar{Q}_{\varphi-}\ ,
\label{Jmassrelmin}
\end{equation}
which are analogous to the relations Eq.~(\ref{Jmassrel}) in the asymptotic region  $\Sigma_+$.
The angular momentum can be obtained from the asymptotic behaviour
of $\bar\omega$,  
\begin{equation}
\bar\omega = - \frac{\tilde{g}_{\bar t \bar \phi}}{\tilde{g}_{\bar \phi \bar \phi}}
= \left(\omega -\omega_{-\infty}\right)\frac{e^{-\gamma/2}}{{\cal A}_-}
\underset{\eta \to - \infty} \longrightarrow \frac{2 J_{\rm J-}}{\bar{\eta}^3} \ ,
\label{barom}
\end{equation}
which yields
\begin{equation}
J_{\rm J-} = J_{\rm E-} e^{-2\gamma}{\cal A}_-^{-2} 
= \bar{J}_{\rm E-} {\cal A}_-^{-2} \ .
\label{JJm}
\end{equation}
Let us reintroduce the Newton constant $G$. 
In the asymptotic region $\eta \to - \infty$
the effective Newton constant is 
$G_{\rm eff}= G_*/F(\Phi_{-\infty}) = G_*{\cal A}_-^2$.
This leads to the simple relation
\begin{equation}
G_{\rm eff}J_{\rm J-} = G_* \bar{J}_{\rm E-}  \ .
\label{JJGm}
\end{equation}

In \cite{Kleihaus:2014dla,Chew:2016epf} several 
relations between the global charges in the Einstein frame
have been found, such as
\begin{equation}
J_{\rm E +} = e^{-2\gamma} J_{\rm E -}    = \bar J_{\rm E -} \ .
\label{j-m-rel1}
\end{equation}
In the Jordan frame these relations then become
\begin{equation}
G_{\rm eff}J_{J-} = G \bar{J}_{E-} = G J_{E+} = G J_{J+} \ .
\end{equation}
Note that the simple relation between masses and angular momentum in 
the Einstein frame \cite{Kleihaus:2014dla,Chew:2016epf},
\begin{equation}
M_{\rm E +} + M_{\rm E -} 
= 2 \omega_{-\infty} e^{-2\gamma} J_{\rm E -} 
= 2 \omega_{-\infty} J_{\rm E +} \ 
\label{j-m-rel}
\end{equation}
has no simple analog in the Jordan frame for the gravitational mass or for the
Schwarzschild mass.

\subsection{Quadrupole Moment}

Let us now turn the derivation of the quadrupole moment ${\cal Q}$ in both frames
in both asymptotic regions.
For  these calculations we employ the definition of the quadrupole moment as given in 
\cite{Hoenselaers:1992bm}.

\subsubsection{Asymptotic region $\Sigma_+$}

The quadrupole moment for rotating wormholes in the Einstein frame 
in the asymptotic region $\Sigma_+$ was derived in \cite{Chew:2016epf},
\begin{equation}
{\cal Q}_{E+} = -f_3 + M_{E+}\eta_0^2 +\frac{M_{E+}\left(M_{E+}^2-D^2\right)}{3} \ ,
\label{quadEp} 
\end{equation}
where $f_3$ is a coefficient appearing in the third order terms of the 
expansion of $f$ in $1/\eta$.

Let us now calculate the quadrupole moment
in the Jordan frame, repeating the suitably modified steps in
\cite{Chew:2016epf}.
To this end we consider a time-like Killing vector field $K$ on the space-time manifold
with metric $\tilde{g}$, and $\tilde{\lambda}$ is the squared norm of $K$. We define the
metric $\tilde{h}$ on a 3-dimensional space by the projection
\begin{equation}
\tilde{h} = -\tilde{\lambda} \tilde{g} + \tilde{K} \otimes \tilde{K} \ ,
\label{metrich}
\end{equation}
where the function $\tilde{\lambda}$ and the 1-form  $\tilde{K}$ in the Jordan frame are
related to $\lambda$ and $K$ in the Einstein frame via
\begin{equation}
\tilde{\lambda} = {\cal A}^2\lambda \ ,  \ \ \ \
\tilde{K}={\cal A}^2 K \ ,
\label{tildlamK}
\end{equation}
respectively.
Consequently, the projected metric in the Jordan frame is related to
the projected metric in the Einstein frame by
\begin{equation}
\tilde{h} = {\cal A}^4 h\ ,  \ 
\label{tildh}
\end{equation}
with 
\begin{equation}
h = \left(1+\frac{r_0^2}{r^2}\right)^2
\left[ e^\nu \left(d\rho^2+dz^2\right) + \rho^2 d\varphi^2\right] \ ,
\label{tildh2}
\end{equation}
where we have introduced quasi-isotropic coordinates, and $r=\sqrt{\rho^2+z^2}$. 
Here we have taken into account that the function $\omega$ does not contribute
to the quadrupole moment.
We note that the function ${\cal A}$ behaves like
\begin{equation}
{\cal A}= 1+ \frac{a_1}{r}+ \frac{a_2}{r^2} + \cdots 
\end{equation}
in the asymptotic region $\Sigma_+$.

We recall that a 3-dimensional space $({\cal M},h)$ is called asymptotically flat if it can
be conformally mapped to a manifold $(\hat{\cal M},\hat{h})$ with
the properties
\begin{itemize}
\item[(i)] $\hat{{\cal M}}= {\cal M}\cup \Lambda $ where $\Lambda \in \hat{{\cal M}}$ 
\item[(ii)] $\left. \Omega \right|_\Lambda = \hat\nabla_i \left. \Omega \right|_\Lambda =0 $
            and 
	    $\hat\nabla_i\hat\nabla_j \left. \Omega \right|_\Lambda =\left. 2\hat{h}_{ij}\right|_\Lambda$, 
	    where $\hat{h}_{ij} =\Omega^2 h_{ij}$ .
\end{itemize}
In \cite{Hoenselaers:1992bm} the complex multipole tensors are defined recursively,
\begin{eqnarray}
\hat{\cal P}^{(0)} & = & \hat{\Phi} \ ,
\nonumber\\
\hat{\cal P}_i^{(1)} & = & \partial_i\hat{\Phi} \ ,
\nonumber\\
\hat{\cal P}_{i_1 \cdots i_{n+1}}^{(n+1)}
& = & {\cal C}\left[ \hat{\nabla}_{i_{n+1}} \hat{\cal P}_{i_1 \cdots i_{n}}^{(n)}
       -\frac{1}{2}n(2n-1)\hat{R}_{i_1 i_2} \hat{\cal P}_{i_3 \cdots i_{n+1}}^{(n-1)}
      \right] \ .
\label{cplxP}      
\end{eqnarray}
Here ${\cal C}$ denotes the symmetric trace-free part, $\hat{R}_{ij}$ is the Ricci tensor,
$\hat{\nabla}_i$ is the covariant derivative on $(\hat{\cal M},\hat{h})$, and 
$\hat{\Phi}=\Omega^{-1/2}\Phi$, where $\Phi=(\tilde\lambda-1/\tilde\lambda)/4$ 
is the complex mass potential.
For $n=1$ we find for the complex quadrupole 
\begin{equation}
\hat{\cal P}_{ij}^{(2)}
 = {\cal C}\left[ \hat{\nabla}_{j}\hat{\nabla}_{i} \hat{\Phi}
       -\frac{1}{2}\hat{R}_{ij} \hat{\Phi}
      \right] \ .
\label{quad1}     
\end{equation}

Next we consider the coordinate transformation
\begin{equation}
\rho' = \frac{\rho}{\rho^2+z^2} \ , \ \ \ \ 
z'    = \frac{z}{\rho^2+z^2} \ ,
\label{transprime}
\end{equation}
which leads to
\begin{equation}
\tilde{h} = \frac{1}{{r'}^4}{\cal A}^4
 \left(1+\frac{{\rho'}^2+{z'}^2}{{r'}_0^2}\right)^2
\left[ e^\nu \left(d{\rho'}^2+d{z'}^2\right) + {\rho'}^2 d\varphi^2\right]
 \ 
\label{dh2t}
\end{equation}
with ${r'}^2 = {\rho'}^2 +{z'}^2$. 

An obvious choice for the conformal factor $\Omega$ seems to be
\begin{equation}
\Omega = {r}'^2{\cal A}^{-2}
 \left(1+\frac{{\rho'}^2+{z'}^2}{{r'}_0^2}\right)^{-1} \ . 
\label{Ome1}
\end{equation}
However, this choice would introduce non-analytic terms in the expressions (\ref{cplxP})
and even a divergent quadrupole moment.
It was noted in \cite{Geroch:1970cd} and \cite{Hansen:1974zz} that the asymptotic flatness condition
does not uniquely determine the conformal factor. In fact the freedom of the choice for the
conformal factor is related to the choice of the origin in $\hat{\cal M}$. The preferred
choice is the centre of mass, where the dipole moment $\hat{\cal P}_i^{(1)}(0)$ vanishes.

Therefore we consider the conformal factor of the form 
\begin{equation}
\Omega' = \Omega (1 +\sigma_1 {r'} +\sigma_2 {r'}^2 + \cdots )^2
\label{Ome2}
\end{equation}
and determine the constants $\sigma_1, \sigma_2, \cdots ,$ such that
$\hat{\cal P}_\rho^{(1)}$ and $\hat{\cal P}_z^{(1)}$ vanish
at ${r'}=0$. This yields
\begin{equation}
\sigma_1 = -\frac{1}{2}\frac{a_1^2-2 M_{E+} a_1 + 2 a_2}{M_{E+}-a_1}  \ ,
\label{sigm1}
\end{equation}
with $\sigma_2$ arbitrary. This choice for the conformal factor also leads to a finite
quadrupole moment.
Using the expansions of the metric functions $f$ and $\nu$ we then find
for the mass and the quadrupole moment in the Jordan frame
\begin{eqnarray}
\mu_{\rm J+} & = & - \hat{\cal P}{(0)} = M_{\rm E +} -a_1 \ , 
\label{mujp}\\
{\cal Q}_{\rm J+} & = & \frac{1}{2}\hat{\cal P}_{zz}^{(2)}(0) = 
-f_3 + \frac{2}{3} M_{\rm E +}\eta_0^2 +\frac{1}{3}M_{\rm E +} c_2 -\frac{1}{3}a_1 c_2
 \ , 
\label{Quad}
\end{eqnarray}
respectively,
with $c_2 = M_{\rm E +}^2 +\eta_0^2-D^2$.
The coefficient $a_1$ may we written conveniently as 
\begin{equation}
a_1= -\left. \frac{d{\cal A}}{d\varphi}\right|_{\varphi=0}Q_\varphi
   =  -\left. \frac{d{\rm ln}{\cal A}}{d\varphi}\right|_{\varphi=0}Q_\varphi \ ,
\label{ceffa1}
\end{equation}
since ${\cal A}(0)=1$.   
We observe that the zeroth multipole moment agrees with the 
gravitional mass, $\mu_{\rm J+} = M_{\rm K+}$.

Comparison with the quadrupole moment in the Einstein frame yields
\begin{equation}
{\cal Q}_{J+} = {\cal Q}_{E+}
+\frac{1}{3}\left. \frac{d{\rm ln}{\cal A}}{d\varphi}\right|_{\varphi=0}Q_\varphi c_2 \ .
\label{Quadrelation}
\end{equation}
Consequently, for STT with  $[d{\cal A}/d\varphi]_{\varphi=0}=0$ 
the masses and quadrupole moments in the asymptotic region $\Sigma_+$ 
conincide in the Jordan and the Einstein frame.
Finally, we give the quadrupole moment in the Jordan frame in terms of the
 gravitational mass and the charges
\begin{equation}
{\cal Q}_{\rm J+} = 
-f_3+\mu_{\rm J+}\eta_0^2 +\frac{1}{3}\mu_{\rm J+}\left(\mu^2_{\rm J+}-Q_\psi^2+Q_\varphi^2\right)
-\frac{\alpha_\infty Q_\varphi}{3}\left(2\mu^2_{\rm J+}+2\eta_0^2-\mu_{\rm J+}\alpha_\infty Q_\varphi \right) \ ,
\label{QuadJ}
\end{equation}
with $\alpha_\infty =\left. \frac{d{\rm ln}{\cal A}}{d\varphi}\right|_{\varphi=0}$.

\subsubsection{Asymptotic region $\Sigma_-$}

Let us now turn to the asymptotic region $\Sigma_-$. Here we first
apply the coordinate transformation Eq.~(\ref{transJ}) to obtain the line element in the Jordan frame,
such that it tends asymptotically to the Minkowski form, 
\begin{equation}
d\tilde{s}^2 = \tilde{\cal A}^2
              \left[
                    -e^{\bar{f}} d\bar{t}^2 + e^{-\bar{f}}
	            \left(e^\nu (d\bar{\eta}^2 + \bar{q}d\theta^2) 
	       	          + \bar{q}\sin^2\theta (d\bar{\phi} -\bar{\omega} d\bar{t})^2
		     \right)
	      \right] \ ,
\label{dtils2}
\end{equation}
where we defined
\begin{equation}
\tilde{\cal A}	= \frac{\cal A}{{\cal A}}_- \ , \ \ \ 
 \bar{f}  = f-\gamma  \ , \ \ \  \bar{q} = \bar{\eta}^2+\bar{\eta}_0^2 \ , \ \ \ 
\bar{\eta}   = e^{-\gamma/2} {\cal A}_{-}\,\eta   \ , \ \ \ 
\bar{\eta}_0 = e^{-\gamma/2} {\cal A}_{-}\,\eta_0 \ , \ \ \ 
 \bar{\omega} = \frac{e^{-\gamma/2}}{{\cal A}_{-}}(\omega-\omega_{-\infty}) \ .
\label{tilfuns}
\end{equation}
Now we can proceed as before. Note however, since the limit $\bar\eta \to -\infty$ corresponds
to the limit $r\to 0$ in isotropic coordinates we do not introduce the coordinates ${\rho'}$,
${z'}$. In the region $\Sigma_-$ the coordinates $\bar\eta$ and $r$ are related by
\begin{equation}
\bar\eta = \frac{1}{r_0}\left(\frac{r}{r_0}-\frac{r_0}{r}\right)
         =-\frac{1}{r}\left(1-\frac{r^2}{r_0^2}\right) \ ,
\label{isocoordsigminus}
\end{equation}
and $\bar\eta_0 = 2/r_0$.

Using the expansions of the metric functions $\bar{f}$ and $\nu$ in the limit $\bar\eta \to -\infty$
the outcome for the mass and the quadrupole moment in the Jordan frame is of the form
\begin{eqnarray}
\mu_{J-} & = & - \hat{\cal P}{(0)} = \bar{M}_{\rm E-}\, -\hat{a}_1 \ , 
\label{mujmin}\\
{\cal Q}_{J-} & = & \frac{1}{2}\hat{\cal P}_{zz}^{(2)}(0) = 
-\hat{f}_3 + \frac{2}{3}\bar{M}_{\rm E-}\,\bar{\eta}_0^2 +\frac{1}{3}\bar{M}_{\rm E-}\, \hat{c}_2 
-\frac{1}{3}\hat{a}_1 \hat{c}_2
 \ , 
\label{Quadmin}
\end{eqnarray}
where the constants can be read off from the asymptotic expansion of the 
metric functions and the gravitational scalar field in terms of $\bar{\eta}^{-1}$,
\begin{equation}
\hat{a}_1 = e^{-\gamma/2}{\cal A}_-\, a_1 \ , \ \ \ 
\hat{f}_3 = e^{-3\gamma/2}{\cal A}_-^3\, f_3 \ , \ \ \ 
\hat{c}_2  = e^{-\gamma}{\cal A}_-^2\, c_2 \ ,
\label{tilcons}
\end{equation}
where $a_1$, $f_3$ and $c_2$ are the coefficients in the expansion
with respect to $\eta^{-1}$.
The coefficient $a_1$ is related to the charge of 
the gravitational scalar field, 
\begin{equation}
a_1= -\left. \frac{d{\rm ln}{\cal A}}{d\varphi}\right|_{\varphi_-}Q_{\varphi -} \ .
\label{a1min}
\end{equation}
Consequently, $\mu_{J-} = M_{\rm K-}$ and 
\begin{equation}
{\cal Q}_{J-} = e^{-3\gamma/2}{\cal A}_-^3 \left\{
-f_3 + \frac{2}{3} M_{\rm E-}\,\eta_0^2 +\frac{1}{3}M_{\rm E-}\, c_2 
     +\frac{1}{3}\left. \frac{d{\rm ln}{\cal A}}{d\varphi}\right|_{\varphi_-}Q_{\varphi -}\, c_2
\right\}\ .
\label{Quadmin1}
\end{equation}
In the Einstein frame the quadrupole moment reads
\begin{equation}
{\cal Q}_{E-} = e^{-3\gamma/2}\left\{
-f_3 + \frac{2}{3} M_{\rm E-}\,\eta_0^2 +\frac{1}{3}M_{\rm E-}\, c_2 
\right\} \ .
\label{QuadEinmin}
\end{equation}
Thus the quadrupole moments in the Jordan frame and the Einstein frame are related 
by
\begin{equation}
{\cal Q}_{J-} = {\cal A}_-^3\left\{{\cal Q}_{E-}
+\frac{1}{3}\left. \frac{d{\rm ln}{\cal A}}{d\varphi}\right|_{\varphi_-}
  \bar{Q}_{\varphi-}\, \bar{c}_2\right\} \ ,
\label{Quadrelationmin}
\end{equation}
with $\bar{c}_2=e^{-\gamma} c_2$.

\subsection{Geometric Properties}

Let us next consider the geometric properties of the wormhole solutions
in scalar-tensor theory. 
Clearly, the geometrical properties of the spacetime depend on
the frame.
We first address the equatorial (or circumferential) radius $\tilde{R}_e$.
Because of the rotation, the throat deforms and its circumference is largest in the
equatorial plane. Therefore a study of $\tilde{R}_e$ reveals the location of the throat.
In the Jordan frame $\tilde{R}_e$ is given by 
\begin{equation}
\label{Re_jf}
\tilde{R}_e = \sqrt{\eta^2+\eta_0^2}\left[{\cal A}(\varphi) e^{-f/2}\right]_{\theta=\pi/2}  
= \frac{\eta_0}{\cos(x)}\left[{\cal A}(\varphi) e^{-f/2}\right]_{\theta=\pi/2} \ ,
\end{equation}
where we defined $\eta = \eta_0 \tan(x)$.
Consequently, the conditions that the throat is located at $x_t$ are given by
\begin{equation}
\frac{d}{dx}\tilde{R}_e(x_t) = 0 \ , \ \ \ {\rm and } \ \ 
\frac{d^2}{dx^2}\tilde{R}_e(x_t) > 0  \ .
\label{thrconds}
\end{equation}

\subsection{Violation of the Null Energy Condition}

Let us finally address the violation of the Null Energy Condition (NEC) 
in both frames.
In the Einstein frame,
we consider the quantity
\begin{equation}
\Xi = R_{\mu\nu} k^\mu k^\nu \ ,
\label{xidef}
\end{equation}
with null vector  \cite{Kashargin:2008pk}
\begin{equation}
k^\mu = \left(e^{-f/2}, e^{f/2-\nu/2}, 0, \omega e^{-f/2}\right) \ .
\label{kmudef}
\end{equation}
Taking into account the Einstein equations and the phantom field equation
we obtain
\begin{equation}
\Xi = - 2 D^2\, \frac{e^{f-\nu}}{q^2}  \ .
\label{xi}
\end{equation}
Since $\Xi$ is non-positive, the NEC is violated everywhere 
\cite{Chew:2016epf}.
%
%In order to obtain a global scale invariant quantity as measure of the NEC violation we
%define for later reference for the symmetric wormholes ???
%
%\begin{equation}
%Y = \frac{1}{R}\int{\Xi \sqrt{-g} d\eta d\theta d\varphi} 
%   =-8\pi \frac{D^2}{R} \int_0^\infty \frac{d\eta}{\eta^2+\eta_0^2}
%   =-4\pi^2 \frac{D^2}{R \eta_0} \ .
%\label{Yxi}
%\end{equation}
%

In the Jordan frame, 
we consider the Einstein equations
\begin{equation}
\tilde{G}_{\mu\nu} = \tilde{T}_{\mu\nu} \ ,
\label{tTdef2}
\end{equation}
where $\tilde{G}_{\mu\nu}$  is the Einstein tensor computed
with the metric $\tilde{g}_{\mu\nu}$, and $\tilde{T}_{\mu\nu}$
is defined by the left hand side of Eq.~(\ref{tTdef2}).
Analogously to Eq.~(\ref{xidef}) we define
\begin{equation}
\tilde \Xi = \tilde R_{\mu\nu} \tilde k^\mu \tilde k^\nu 
=\tilde{T}_{\mu\nu} \tilde k^\mu \tilde k^\nu \ ,
\label{xidef2}
\end{equation}
with null vector  $\tilde k^\mu = k^\mu$.
%
%\begin{equation}
%\tilde k^\mu =  \left(e^{-f/2}, e^{f/2-\nu/2}, 0, \omega e^{-f/2}\right) \ .
%\label{kmudef2}
%\end{equation}
This yields
\begin{equation}
\tilde \Xi = - e^{f-\nu} \left( 
2 \frac{D^2}{q^2}
-\partial_\eta \nu \frac{\partial_\eta {\cal A}}{{\cal A}}
+2\frac{\partial_{\eta\eta} {\cal A}}{{\cal A}}
-\left(2\frac{\partial_{\eta} {\cal A}}{{\cal A}}\right)^2
\right) \ .
\label{xi2}
\end{equation}
%
%%We note that in the Einstein frame the NEC is violated everywhere,
%%since $\Xi$ is non-positive. However, 
%Thus in the Jordan frame the NEC need not be violated everywhere.
%%this need not to be true.

In order to demonstrate the violation of the NEC we consider
the densities
\begin{eqnarray} 
\Xi' & = & \sqrt{-g} \Xi = - 2 D^2\, \frac{\sin\theta}{q} \  ,
\label{Xip}\\
\tilde{\Xi}' & = & \sqrt{-\tilde{g}} \tilde{\Xi}
 = 
-{\cal A}^4 \frac{\sin\theta}{q}
\left( 
2 D^2 
-q\partial_\eta \nu \frac{q\partial_\eta {\cal A}}{{\cal A}}
+2  \frac{q^2\partial_{\eta\eta} {\cal A}}{{\cal A}}
-\left(2 \frac{q\partial_{\eta} {\cal A}}{{\cal A}}\right)^2
\right) \ .
\label{tXip}
\end{eqnarray} 
In the Einstein frame $\Xi'$ does not involve the metric functions
(except for the auxiliary function $q$). Thus  the violation of 
the NEC is the same for all solutions with the same value of $D$.
In constrast, in the Jordan frame $\tilde{\Xi}'$ depends 
on the metric function $\nu$ and on the gravitational scalar field
$\varphi$.
%It can be easily seen that the NEC is violated in the Jordan frame as well.
%For any regular solution the conformal factor ${\cal A}$ remains finite, 
%implying that $\partial_\eta{\cal A}$ tends to zero in $\Sigma_\pm$. 
%Consequently, the densities $\Xi'$ and $\tilde{\Xi}'$
%then coincide in $\Sigma_\pm$ and are non-positive. Hence the NEC is violated 
%at least in  $\Sigma_\pm$.

\section{Wormholes in Three Examples of Scalar Tensor Theories}

In the following we will consider three examples 
of STT, which we specify by their respective coupling function
${\cal A}(\varphi)$: 
%STT-1 with ${\cal A}_1(\varphi)=e^{\beta\varphi^2/2}$, 
%STT-2 with ${\cal A}_2(\varphi)=e^{\alpha\varphi}$ and
%STT-3 with  ${\cal A}_3(\varphi)=\cosh(\varphi/\sqrt{3})$.
\begin{equation}
{\cal A}_1(\varphi)=e^{\beta\varphi^2/2} 
\ \ \text{(STT-1)} \  , \ \ \ 
{\cal A}_2(\varphi)=e^{\alpha\varphi}
\ \ \text{(STT-2)} \  , \ \ \ 
{\cal A}_3(\varphi)=\cosh(\varphi/\sqrt{3})
\ \ \text{(STT-3)} \ . 
\label{cala_examples}
\end{equation}
The first example has been employed by Damour and Esposito-Farese,
when demonstrating 
the scalarization of neutron  stars \cite{Damour:1993hw,Damour:1996ke}.
The second example represents Brans-Dicke theory
\cite{Brans:1961sx},
while the third example appears to be new and has been inspired by \cite{Barcelo:1999hq}.
We note, that
the rotating wormhole solutions of GR \cite{Kleihaus:2014dla,Chew:2016epf}
are also solutions of the STT equations
in the first and the third example. However, this is not the case
in the second example.

In all cases the boundary condition for the gravitational scalar field $\varphi$
in the asymptotic region $\Sigma_+$  is chosen such that it vanishes there, $\varphi(+\infty)=0$. 
Therefore in all examples the coupling function tends to
${\cal A}(0)=1$ in $\Sigma_+$. In contrast, in the asymptotic region $\Sigma_-$
the gravitational scalar field 
$\varphi$ assumes a finite value $\varphi_-$,  yielding
\begin{equation}
\left. \frac{d{\rm ln}{\cal A}_1}{d\varphi}\right|_{\varphi_-}  = 
\beta \varphi_- \ , \ \ \ \ 
%{\rm STT-1}\  , \ \ \ 
\left. \frac{d{\rm ln}{\cal A}_2}{d\varphi}\right|_{\varphi_-}  = 
\alpha \ , \ \ \ \  
%{\rm STT-2} \  ,  \ \ \ 
\left. \frac{d{\rm ln}{\cal A}_3}{d\varphi}\right|_{\varphi_-}  = 
\sqrt{3}\tanh\left(\frac{\varphi_-}{\sqrt{3}}\right) 
%\ \ {\rm STT-3} \ .
\ .
\label{a1_examples}
\end{equation}
%

%In the first and third example ${\cal A}(\varphi) = 1 +{\cal O}(\varphi^2)$. 
%Here the gravitational mass and the Schwarzschild mass coincide in the Jordan frame 
%and also coincide with the mass in the Einstein frame.
%In the second example $\left.\frac{d{\cal A}}{d\varphi}\right|_{\varphi=0}=\alpha$.
%In this case the gravitational mass and the Schwarzschild mass 
%differ from the tensor mass by $\alpha Q_\varphi$ and $-\alpha Q_\varphi$, respectively.

\subsection{Model STT-1}

We first consider the model STT-1
employed by Damour and Esposito-Farese \cite{Damour:1993hw}
with the non-miminal coupling 
\begin{equation}
{\cal A}= e^{\frac{\beta}{2}\varphi^2} \ .
\label{calA}	 
\end{equation}
With this STT they discovered the phenomenon
of spontaneous scalarization in neutron stars,
when choosing the coupling parameter $\beta$ 
below a critical negative value \cite{Damour:1993hw}.
On the other hand, observations 
on the binary pulsar PSR J1738+0333
impose a lower bound on $\beta$,
$\beta > -4.5$ \cite{Freire:2012mg},
leaving only a small interval of $\beta$
to achieve scalarization 
for a massless gravitational scalar field in neutron stars
\cite{Freire:2012mg,Doneva:2013qva},
but being much less restrictive in the massive case
\cite{Yazadjiev:2016pcb,Doneva:2016xmf}.
We note that for boson stars and hairy black holes
the upper bound on $\beta$ is similar to the one for neutron stars
\cite{Whinnett:1999sc,Alcubierre:2010ea,Ruiz:2012jt,Kleihaus:2015iea}.

In neutron stars and boson stars the phenomenon of spontaneous scalarization
is restricted to negative values of the coupling $\beta$.
Here we show, that scalarization of wormholes arises for arbitrary (finite) values of $\beta$.
This includes, in particular, positive values of $\beta$.
For positive $\beta$ the solutions develop an oscillating gravitational scalar field,
which can give rise to wormholes with a multitude of throats 
and equators in the Jordan frame.
%In the following we start with the negative $\beta$ solutions, 
%and address at the ends of the respective subsections also the domain  of existence
%and the physical properties of the solutions  with positive $\beta$.
In the following subsections 
we present the solutions for general $\beta$,
discuss the domain of existence for negative $\beta$
and subsequently for positive $\beta$,
and then analyze the physical properties of these solutions.

\subsubsection{Solutions}

In STT-1 Eq.~(\ref{redequphi}) for the gravitational scalar field becomes
\begin{equation}
\partial_\eta \left(\hq \partial_\eta \varphi\right) = 
-\beta \frac{\varphi}{\hq}e^{-\beta\varphi^2} Q_\psi^2 \ .
\label{redequphi1}
\end{equation}
Clearly, $\varphi =0$ is always a solution. Hence any wormhole solution of 
Einstein gravity is also a solution of STT-1, 
although with a trivial gravitational scalar field.

However, there are in addition solutions with a non-trivial 
gravitational scalar field, i.e., scalarized wormhole solutions,
which can be obtained from Eq.~(\ref{equphi1}), which now reads
\begin{equation}
\partial_{x} \varphi
= \pm \frac{Q_\psi}{\eta_0} e^{-\frac{\beta}{2}\varphi^2}
\sqrt{1-a^2 e^{\beta\varphi^2}}  \ ,
\label{equphi3}
\end{equation}
with $a^2=  D^2/Q_\psi^2$.
These solutions are obtained numerically.

\subsubsection{Domain of existence}

We start our discussion of the domain of existence
with the case of negative $\beta$.
%respectively positive $\beta$ separately.
Here it is convenient to introduce
the scaled functions and charges
\begin{equation}
\hat{\varphi} = \sqrt{-\beta}\varphi \ , \ \ \ 
\hat{Q}_\psi = \sqrt{-\beta}\frac{Q_\psi}{\eta_0}  \ , \ \ \ 
\hat{D}      = \sqrt{-\beta}\frac{D}{\eta_0}  \ .
\label{bardef1}
\end{equation}
In terms of these quantities Eq.~(\ref{equphi3}) reads
\begin{equation}
\partial_{x} \hat{\varphi} 
= \pm\hat{Q}_\psi e^{\frac{1}{2}\hat{\varphi}^2}
\sqrt{1-a^2 e^{-\hat{\varphi}^2}}  \ .
\label{equphi3sc}
\end{equation}
We note from the ODE Eq.~(\ref{equphi3sc}) that 
$\partial_{x} \hat{\varphi}\neq 0$ 
since  $a^2 e^{-\hat{\varphi}^2} < 1$. Therefore $\hat{\varphi}(x)$ is a 
monotonic function. Consequently, we can rewrite this equation in
integral form
\begin{equation}
\int_0^{\hat{\varphi}(x)} 
e^{-\frac{1}{2}\hat{\varphi}^{'2}}\left\{1-a^2  e^{-\hat{\varphi}^{'2}} \right\}^{-\frac{1}{2}}
d\hat{\varphi}'
= \pm\hat{Q}_\psi\left(x-\frac{\pi}{2}\right) \ .
\label{intequphi}
\end{equation}
To obtain a regular solution we demand that $\hat{\varphi}(x)$ is finite
on the interval $-\infty \leq \eta \leq \infty$, corresponding to
$-\frac{\pi}{2} 
\leq x \leq 
\frac{\pi}{2}$.
Thus in the limiting case, $\hat{\varphi} \to \mp \infty$ as 
$ x \to -\frac{\pi}{2}$, we find a critical value
$\hat{Q}_\psi^{\rm cr}$ for the scalar charge,
\begin{equation}
\hat{Q}_\psi^{\rm cr}   =  
\pm \frac{1}{\pi}\int_{0}^{\infty} 
e^{-\frac{1}{2}\hat{\varphi}^2}\left\{1-a^2  e^{-\hat{\varphi}^2} \right\}^{-\frac{1}{2}}
d\hat{\varphi}
\label{eqQpsicrx1} \ .
\end{equation}
An analytical expression for the critical phantom scalar charge $\hat{Q}_\psi^{\rm cr}$ is given in the Appendix.

\begin{figure}[t!]
\begin{center}
\mbox{(a)
\includegraphics[height=.23\textheight, angle =0]{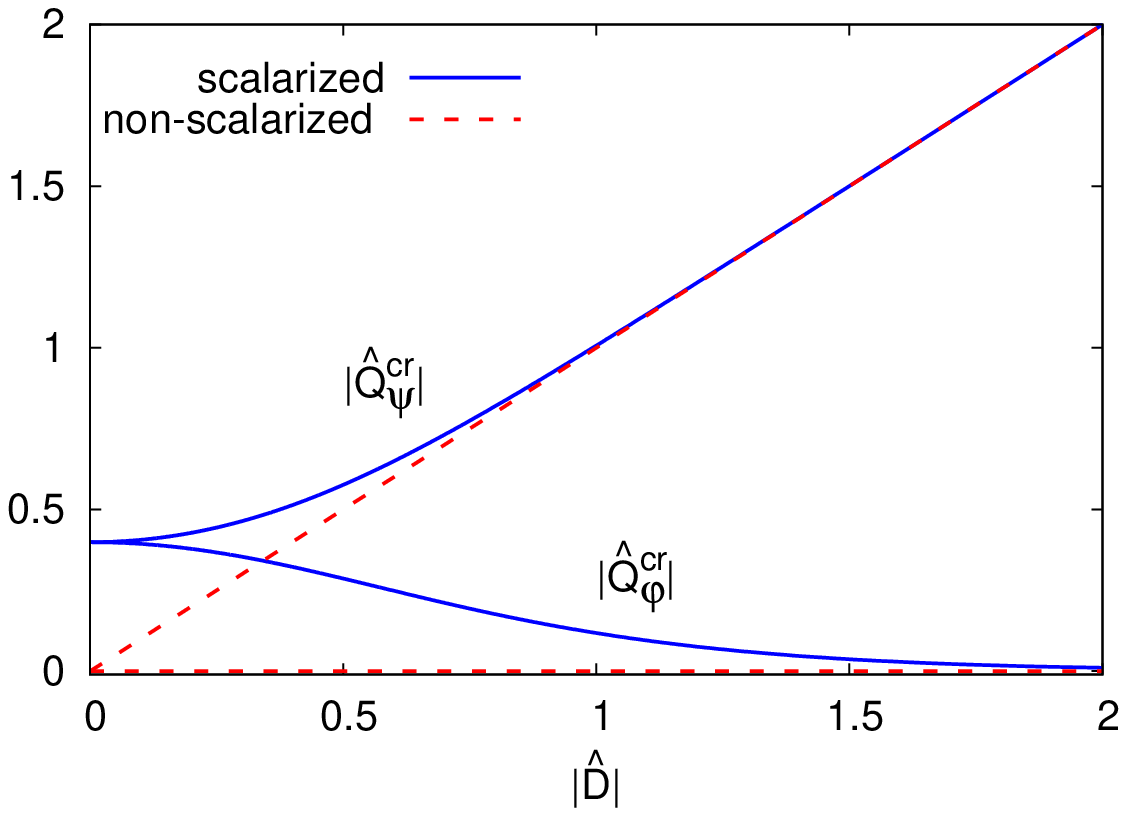}
(b)
\includegraphics[height=.23\textheight, angle =0]{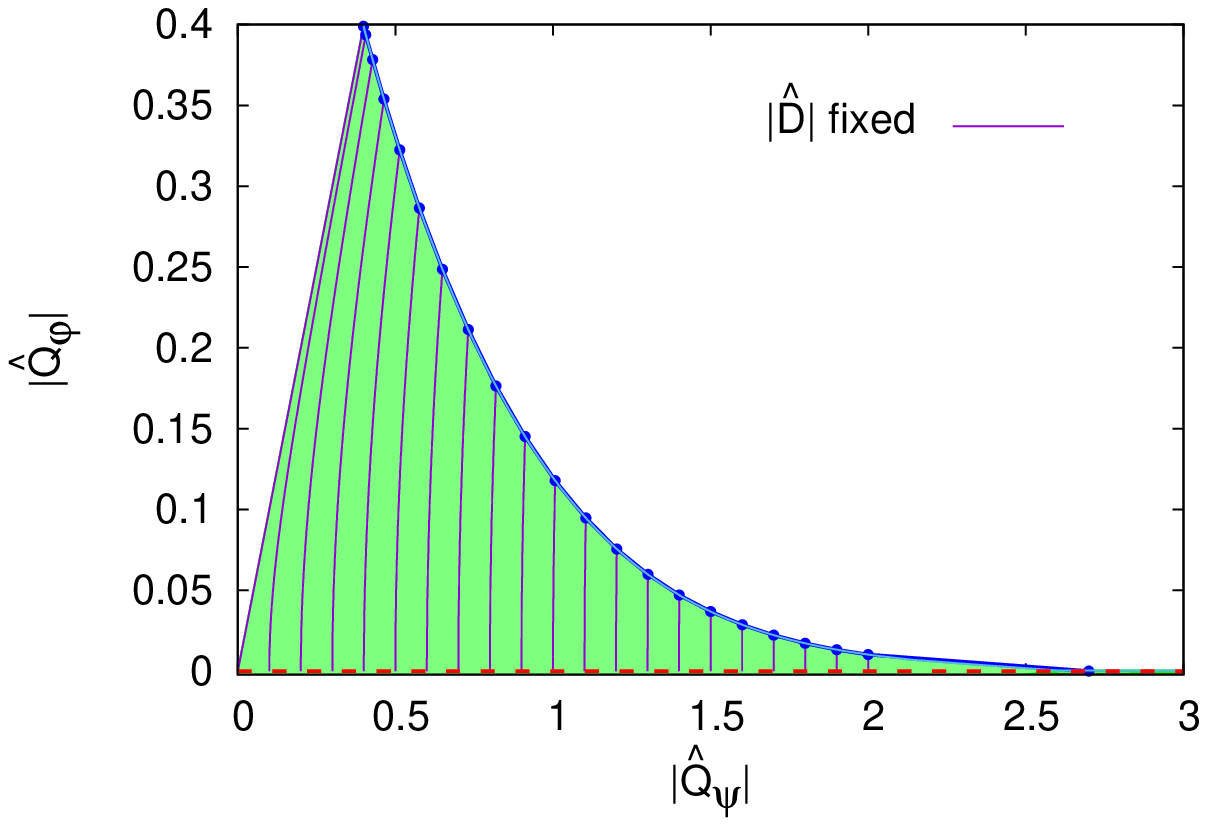}}
\end{center}
\vspace{-0.5cm}
\caption{Wormholes in STT-1 for negative $\beta$:
(a) The scaled critical charges $\hat{Q}_\psi^{\rm cr}$ (upper)
and $\hat{Q}_\varphi^{\rm cr}$ (lower)
versus the scaled quantity $\hat{D}$ with and without
scalarization.
(b)
The domain of existence in the $\hat{Q}_{\psi}$-$\hat{Q}_{\varphi}$ plane.
The lilac curves correspond to fixed values of $\hat{D}$,
which can be read off at the intersections with the abscissa $|\hat{Q}_{\psi}|$.
\label{Fig1}
}
\end{figure}

The corresponding bounds for the gravitational scalar charge 
$\hat{Q}_\varphi^{\rm cr} = \frac{\sqrt{-\beta}}{\eta_0} Q_\varphi^{\rm cr}$ 
can be found from Eq.~(\ref{Qrelation}), 
i.e.~$(\hat{Q}_\varphi^{{\rm cr}})^2=(\hat{Q}_\psi^{{\rm cr}})^2-\hat{D}^2$. 
The bounds $\hat{Q}_\psi^{\rm cr}$ and $\hat{Q}_\varphi^{\rm cr}$ are shown in 
Fig.~\ref{Fig1}(a) as functions of $\hat{D}$.
We observe %from Fig.~\ref{Fig2} that
that with increasing $|\hat{D}|$ the phantom scalar charge
$|\hat{Q}_\psi^{\rm cr}|$ tends rapidly to $|\hat D|$,
while the gravitational scalar charge $|\hat{Q}_\varphi^{\rm cr}|$ tends to zero.

For $\hat{D}=0$, the scalar charges are equal. Here the bound is given by
$|\hat{Q}_\psi^{\rm cr}(0)|= |\hat{Q}_\varphi^{\rm cr}(0)|=1/\sqrt{2\pi}$.
However, the case $\hat{D}=0$ is special, since the contributions 
of the gravitational scalar field and the phantom scalar field in the
stress-energy tensor cancel exactly. Hence the spacetime metric
corresponds to the metric of a Kerr black hole.
In fact, the only acceptable $\hat{D}=0$ solution 
carries no scalar fields at all,
i.e., $\hat{Q}_\psi^{\rm cr}=\hat{Q}_\varphi^{\rm cr}=0$,
since otherwise the scalar fields would diverge at the horizon.

In Fig.~\ref{Fig1}(b) we exhibit the domain of existence in the 
$\hat{Q}_{\psi}$-$\hat{Q}_{\varphi}$ plane.
For each value of $\hat{D}$ we determine the 
bound $\hat{Q}_\psi^{\rm cr}$ from Eq.~(\ref{eqQpsimcr4})
and show $\hat{Q}_\varphi^{\rm cr}$ as a function of 
$\hat{Q}_\psi$ in the interval $\left[0 ,\hat{Q}_\psi^{\rm cr}\right]$. 
In the figure the values of $\hat{D}$ can be read off at the 
intersections of the lilac curves with the abscissa. 

In fact, the abscissa corresponds
to the family of non-scalarized wormhole solutions. 
Thus we see that for each value of $\hat D$ 
a branch of scalarized wormhole solutions 
emerges from the non-scalarized ones. For a given $D$, all wormhole solutions
possess the same mass and the same angular momentum. 
The branches end, when the corresponding critical value 
of $\hat{Q}_\varphi$ is reached.

\begin{figure}[t!]
\begin{center}
\mbox{
(a)
\includegraphics[height=.23\textheight, angle =0]{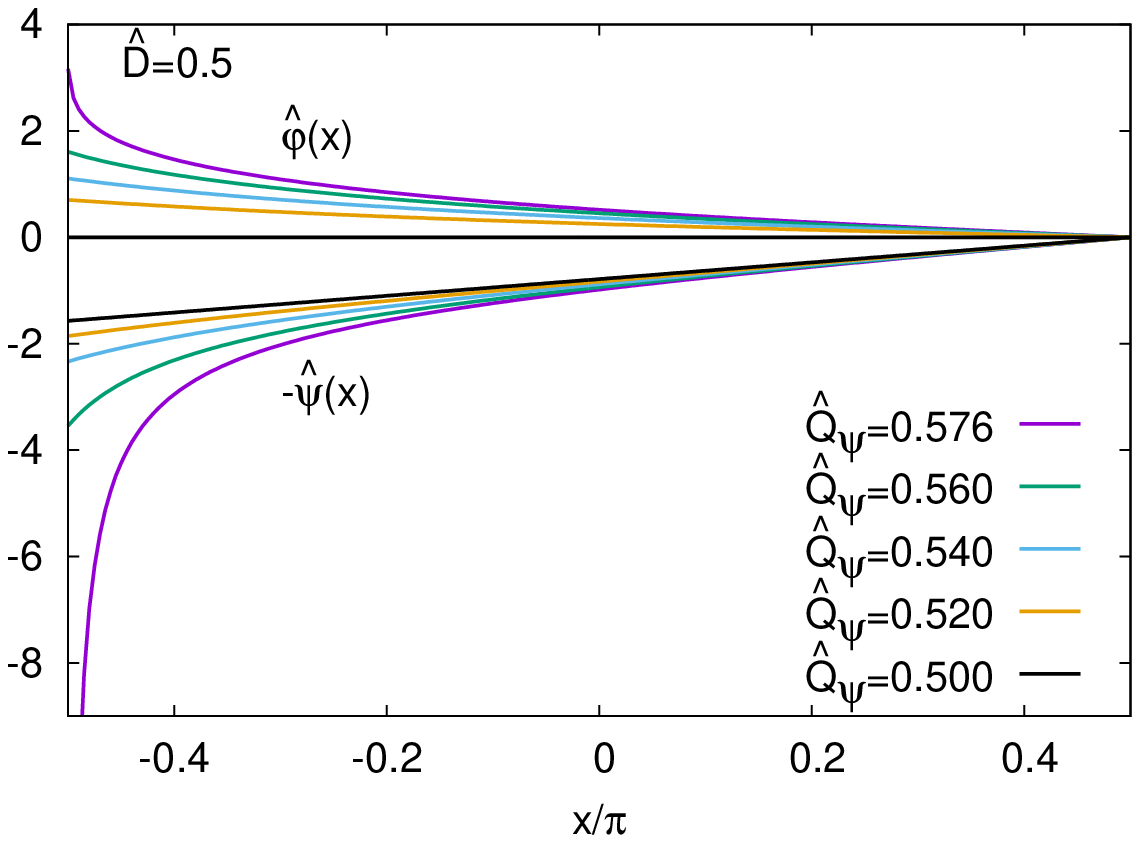}
(b)
\includegraphics[height=.23\textheight, angle =0]{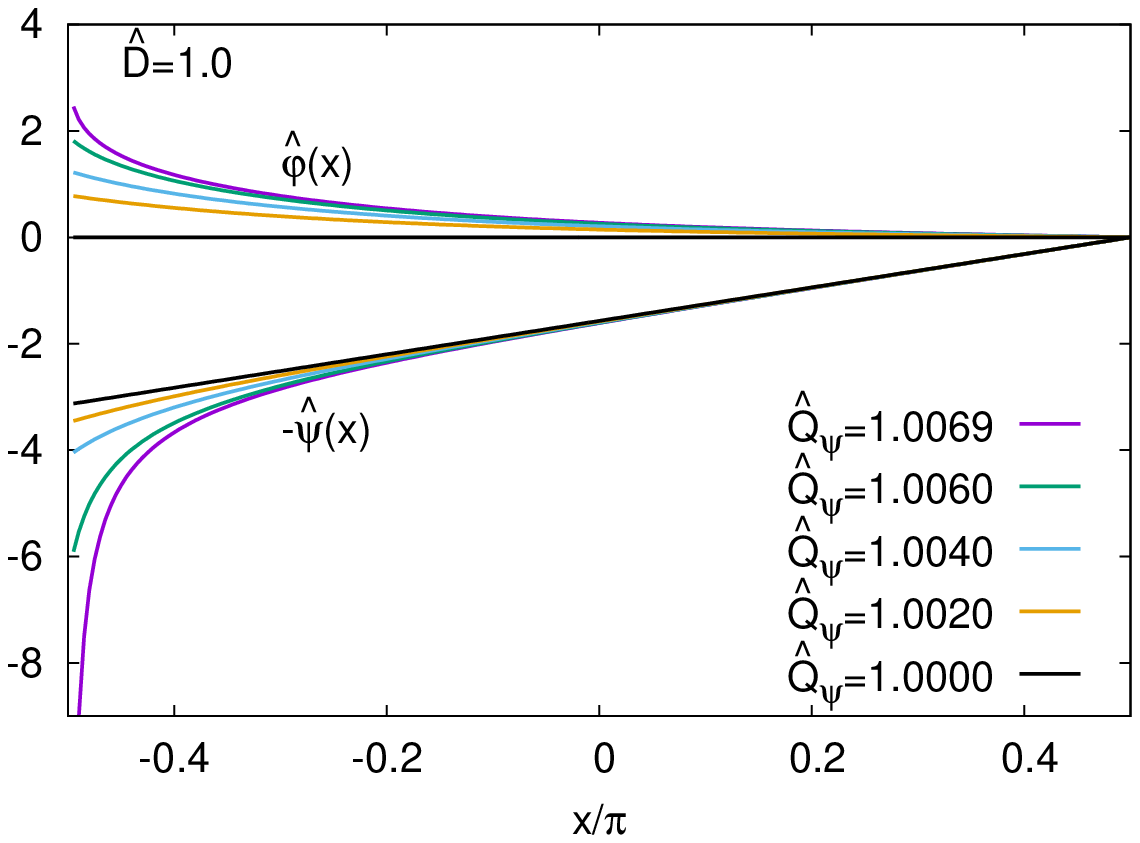}}
\mbox{
(c)
\includegraphics[height=.23\textheight, angle =0]{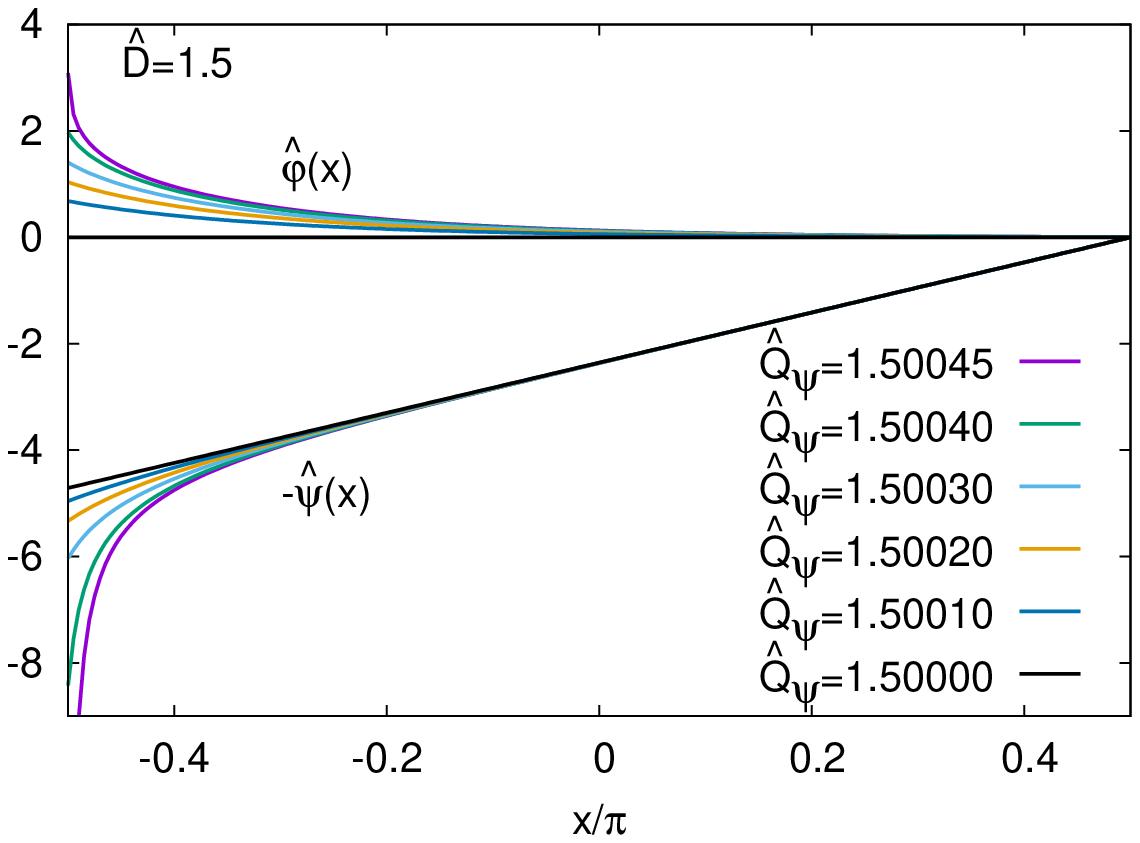}
}
\end{center}
\vspace{-0.5cm}
\caption{Wormholes in STT-1 for negative $\beta$:
The scaled functions $\hat{\varphi}(x)$
and $\hat{\psi}(x)$ (a) for $\hat{D} = 0.5$, (b) for  $\hat{D} =1$
and (c) for $\hat{D} =1.5$
and several values of $\hat{Q}_\psi$.
\label{Fig2}
}
\end{figure}

We exhibit in Fig.~\ref{Fig2} examples of solutions with negative $\beta$.
%As examples for solutions with negative $\beta$ we show in Fig.~\ref{Fig2} 
Here
the scaled functions $\hat{\varphi}$ and 
$\hat{\psi}$ are shown versus the scaled coordinate $x/\pi$ 
for $\hat{D} = 0.5$ (a), $\hat{D} =1$ (b) and $\hat{D} =1.5$ (c),
and several values of
$\hat{Q}_\psi$ ranging from  $\hat{D}$ up to
a value close to the critical value $\hat{Q}_\psi^{\rm cr}$,
Eq.~(\ref{eqQpsicrx1}).

We note that the solution for the phantom field
is the same in the Jordan frame and in the Einstein frame.

Let us next consider the case of positive $\beta$. 
Here we rewrite the first order ODE
as
\begin{equation}
\partial_{x} \varphi 
= \pm \frac{Q_\psi}{\eta_0}
\sqrt{e^{-\beta\varphi^2}-a^2}  \ .
\label{equphi4}
\end{equation}
From the boundary condition $\varphi(\pi/2)=0$ we find 
$\partial_{x} \varphi(\pi/2)=\sqrt{1-a^2}Q_\psi/\eta_0=Q_\varphi/\eta_0$.
Let us suppose $Q_\varphi>0$, then $\varphi$ decreases with decreasing $x$.
If $\varphi$ assumes the value $-\varphi_{\rm ex} = -\sqrt{-2{\rm ln}(a)/\beta}$ at
some point $x_{\rm ex}$, then $\partial_{x} \varphi(x_{\rm ex})=0$ corresponding
to a minimum. 
To continue to $x\leq x_{\rm ex}$ smoothly we have to choose
the lower sign in Eq.~(\ref{equphi4}). 
Thus  $\varphi$ increases with further decreasing $x$, until it reaches 
the value $\varphi_{\rm ex}$, which is a maximum, and so on, until the
boundary $x=-\pi/2$ is reached. 
In this way we have established a bound for the solutions, 
$-\varphi_{\rm ex}\leq \varphi(x)\leq \varphi_{\rm ex}$.
Consequently, solutions exist for all (finite) 
$Q_\psi$, $-1< a< 1$ and $\beta >0$.
Note that also a bound for the derivative $\partial_{x}\varphi$ exists,
$-|Q_\varphi/\eta_0| \leq \partial_{x} \varphi \leq |Q_\varphi/\eta_0|$.

%\subsubsection{Mass, angular momentum and quadrupole moment}
\subsubsection{Physical properties}

We now consider the mass, the angular momentum and the quadrupole moment
in STT-1. 
As discusssed in Section {2.5} the angular momentum 
is the same in both frames.
In the asymptotic region $\Sigma_+$ the coupling function ${\cal A}$ 
satisfies $d{\rm ln}{\cal A}/d\varphi = 0$, and thus
the gravitational mass and the Schwarzschild mass 
in the Jordan frame are the same as the ADM mass in the Einstein frame,
\begin{equation}
M_{\rm K+} = M_{\rm S+} = M_{\rm E+} \ .
\label{D-STTmass}
\end{equation}

In the static case there is a simple relation between the mass in the 
Einstein frame and the scalar charges,
\begin{equation}
M^2_{\rm E+} = Q_\psi^2 - Q_\varphi^2 -\eta_0^2 \ .
\label{mass-charge-rel}
\end{equation}
Consequently, 
\begin{equation}
M^2_{\rm K+} = M^2_{\rm S+} = Q_\psi^2 - Q_\varphi^2 -\eta_0^2 
\label{M2scal}
\end{equation}
for the scalarized static wormholes in STT. 
Since for the non-scalarized wormholes $Q_\varphi =0$, 
the comparision of the masses of scalarized and non-scalarized
wormholes with the same phantom scalar charge yields
\begin{equation}
M^2_{\rm scal} - M^2_{\rm non-scal} = - Q_\varphi^2 \ .
\label{delM2scal}
\end{equation}
Hence for a fixed phantom scalar charge the static scalarized wormholes possess less mass than the
non-scalarized wormholes.

The quadrupole moments are the same in the Jordan frame and in the
Einstein frame. This equivalence arises 
again because $d{\rm ln}{\cal A}/d\varphi = 0$ holds in the
asymptotic region $\Sigma_+$.

In the asymptotic region $\Sigma_-$ the gravitational scalar field assumes 
a finite value $\varphi_-$. Consequently, the masses in the Jordan frame and in
the Einstein frame differ,
\begin{equation}
M_{\rm K-} + M_{\rm S-} = 2 e^{\frac{\beta}{2}\varphi_-^2}  \bar{M}_{\rm E-} \ , \ \ \ 
M_{\rm K-} - M_{\rm S-} = 2 \beta\varphi_-
\sqrt{1-a^2 e^{\beta\varphi_-^2}} \bar{Q}_\psi \ .
\label{D-STTmassmin}
\end{equation}
The angular momentum in the Jordan frame differs from the angular momentum in
the Einstein frame by a factor,
\begin{equation}
%J_{\rm J-}=\bar{J}_{\rm E-} e^{\beta\varphi_-^2} \ .
J_{\rm J-}=\bar{J}_{\rm E-} e^{-\beta\varphi_-^2} \ .
\label{D-STTangmin}
\end{equation}
For the quadrupole moment we find from the general expression, 
Eq.~(\ref{Quadrelationmin}),
\begin{equation}
{\cal Q}_{\rm J-} = e^{\frac{3\beta}{2} \varphi_-^2} {\cal Q}_{E-}
+\frac{1}{3}\beta\varphi_-  e^{\beta\varphi_-^2}
\sqrt{1-a^2 e^{\beta\varphi_-^2}} \bar{Q}_\psi \bar{c}_2 \ .
\label{D-STTquadmin}
\end{equation}

Let us now consider the geometric properties 
of the wormholes in the Jordan frame.
We begin with the location and the equatorial radius of the throat. 
In the Jordan frame
the circumferential radius is defined in Eq.~(\ref{Re_jf}),
and the condition for
the throat coordinate is given in Eq.~(\ref{thrconds}). 
In STT-1 they read
\begin{equation}
\tilde{R}_e(x) = 
\frac{\eta_0}{\cos x} \left. e^{-(-\beta\varphi^2+f)/2} \right|_{\theta=\pi/2} \ , \ \ \ \
\left(\tan x+ \beta\varphi \partial_x \varphi-\frac{1}{2}\partial_x f
\right)_{x_t, \theta=\pi/2} =0 \ .
\label{D-STTthroat}
\end{equation}

\begin{figure}[t!]
\begin{center}
\mbox{
\includegraphics[height=.23\textheight, angle =0]{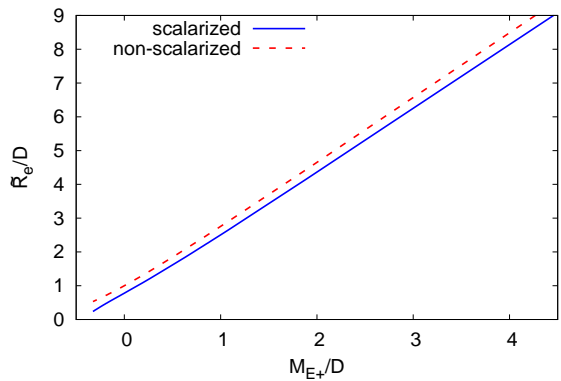}
}
\end{center}
\vspace{-0.5cm}
\caption{
Wormholes in STT-1 for negative values of $\beta$:
The scaled circumferential radius $\tilde{R}_e/D$ of the throat 
of scalarized and non-scalarized wormholes
versus the scaled mass $M_{{\rm E}+}/D$ 
for $D=1$, $Q_\psi = Q_\psi^{\rm cr}$ and 
$\beta= -0.1$.
\label{Fig3}
}
\end{figure}

Starting again with the case of negative $\beta$,
we note that  the geometry of the throat of the wormholes
in the Jordan frame is almost identical to the one in the Einstein
frame for this value of $\beta$. 
The smallness of the deviation is
demonstrated for $\beta=-0.1$ in Fig.~\ref{Fig3}, where we show
the circumferential radius of the throat in the equatorial plane as a function of 
the mass.
Here we set $Q_\psi = Q_\psi^{\rm cr}$ and choose a sequence of rotating wormhole
solutions with fixed $D=1$. Also shown is the throat radius of the non-scalarized
wormhole solutions.
The two curves represent the limits of maximal, respectively vanishing scalarization
for fixed  $D=1$.
%Starting again with the case of negative $\beta$,
%we note that for the observational limit $\beta = -4.5$ the gravitational
%scalar field $\varphi$ is bounded by $|\varphi| \leq 0.1283$,
%yielding for the conformal factor $0.9286 \leq {\cal A}^2 \leq 1$. 
%Consequently, the geometry of the throat of the wormholes
%in the Jordan frame is almost identical to the one in the Einstein
%frame for this value of $\beta$. 
%%Even for $\beta = -1$ there arises no considerable difference.
%%Only for rather large values of $\beta$  
%The smallness of the deviation is
%demonstrated in Fig.~\ref{Fig3} for $\beta=-0.1$.

\begin{figure}[t!]
\begin{center}
\mbox{(a)
\includegraphics[height=.23\textheight, angle =0]{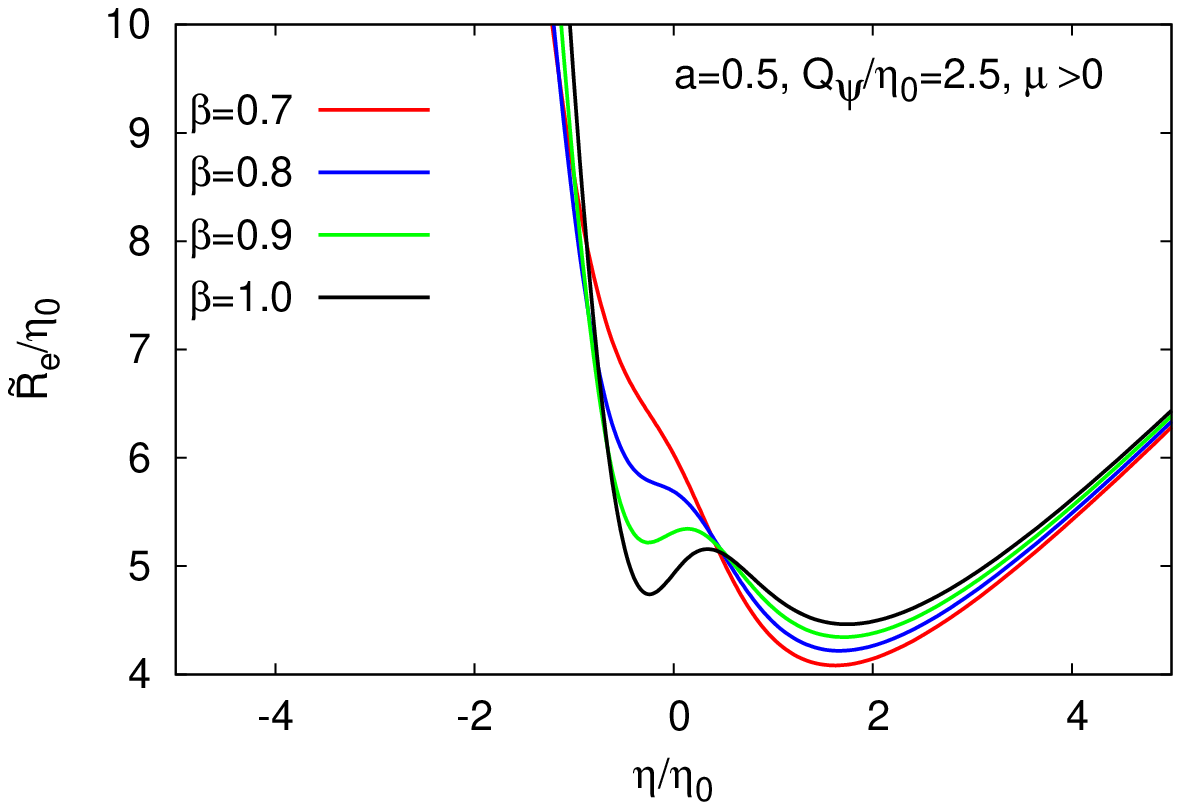}
(b)
\includegraphics[height=.23\textheight, angle =0]{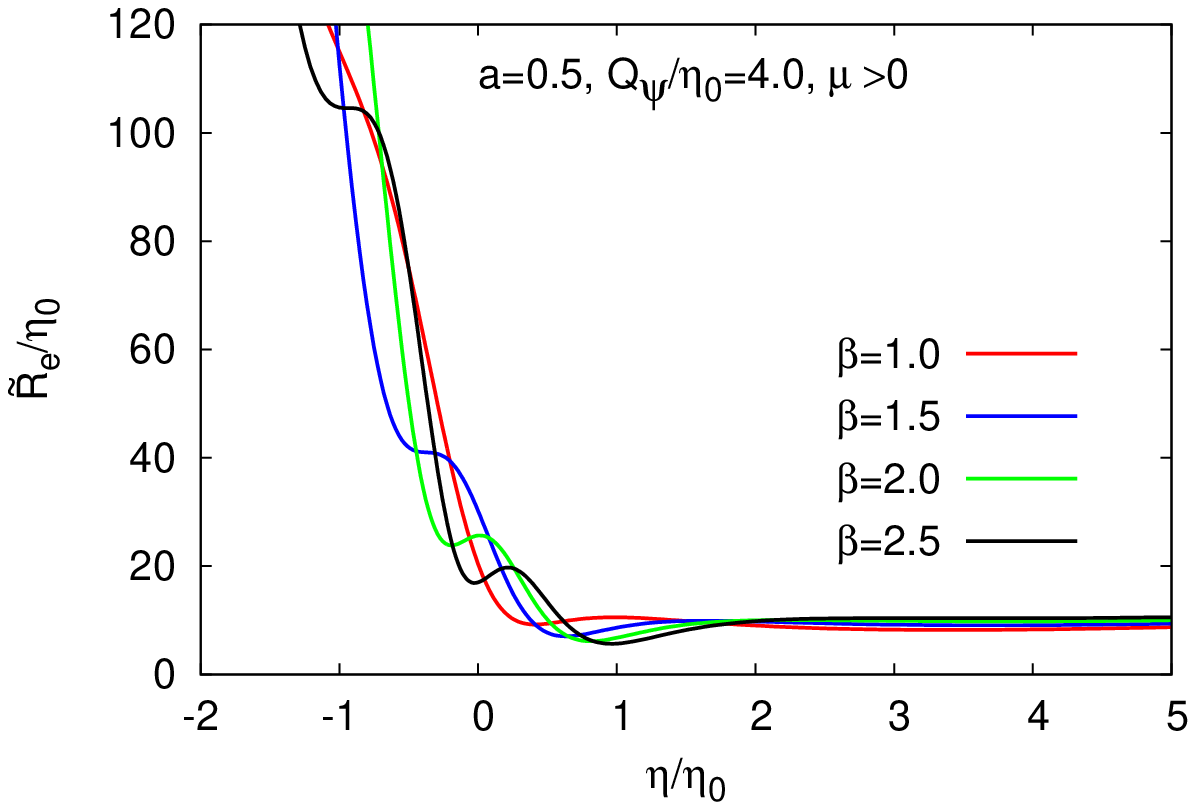}
}
\mbox{(c)
\includegraphics[height=.23\textheight, angle =0]{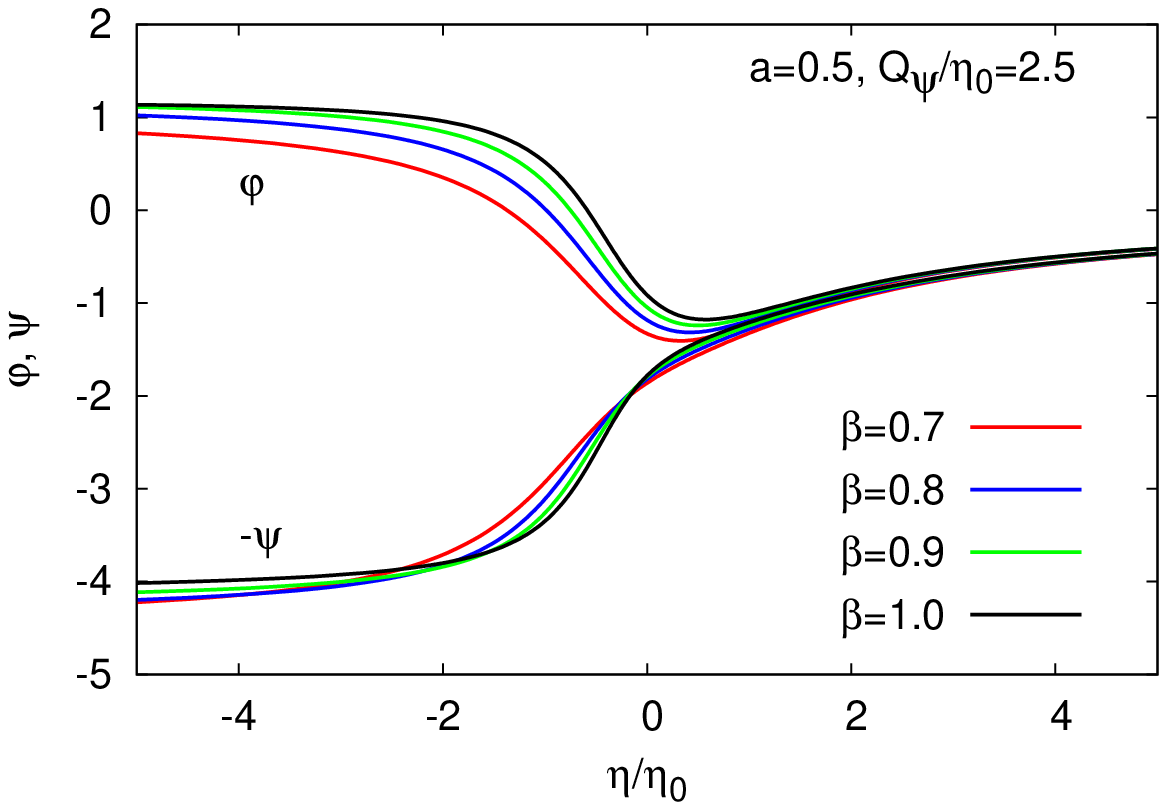}
(d)
\includegraphics[height=.23\textheight, angle =0]{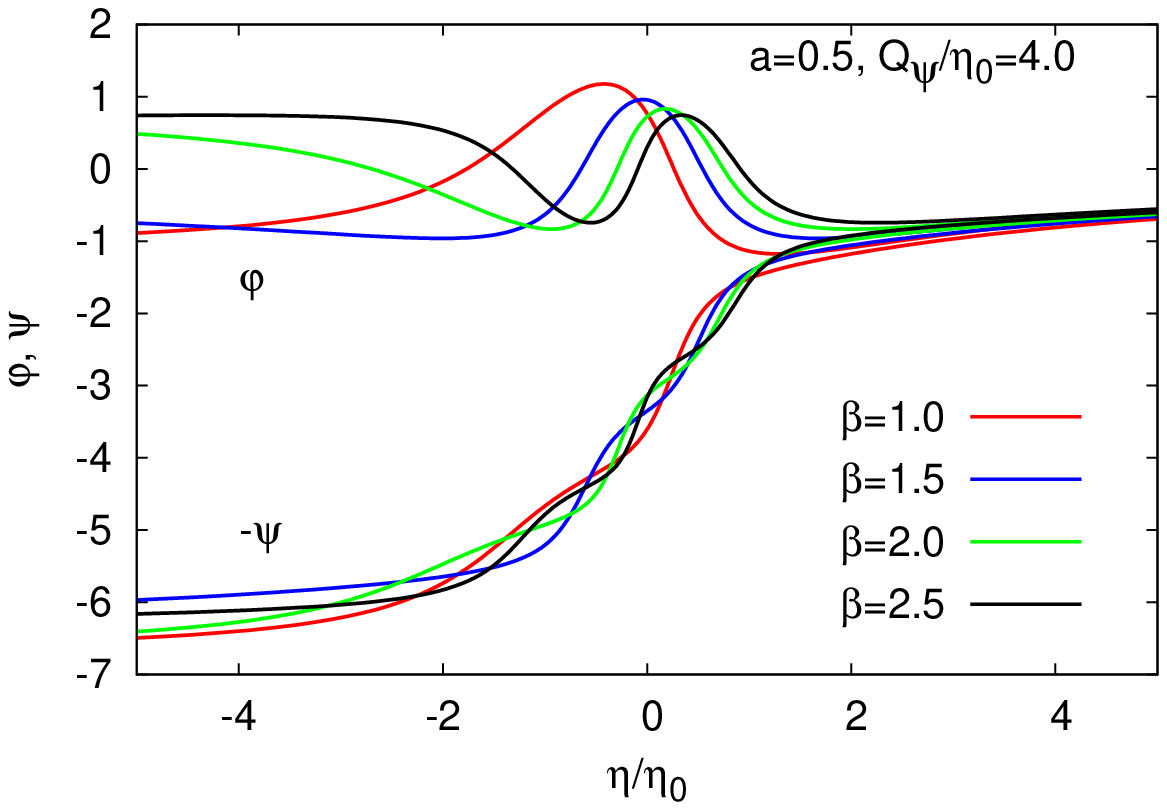}
}
\end{center}
\vspace{-0.5cm}
\caption{
Static wormholes in STT-1 for positive values of $\beta$:
Top: The circumferential radius coordinate $\tilde{R}_e$ 
versus  the radial coordinate $\eta$ 
for several values of $\beta$, and 
$Q_\psi/\eta_0= 2.5$, $a=0.5$ (a) and $Q_\psi/\eta_0= 4.0$, $a=0.5$ (b).
Bottom: The scalar fields $\varphi$ and $\psi$ are shown versus the
coordinate $x$ for several values of $\beta$, and 
$Q_\psi/\eta_0= 2.5$, $a=0.5$ (c) and $Q_\psi/\eta_0= 4.0$, $a=0.5$ (d).
\label{Fig4}
}
\end{figure}

\begin{figure}[t!]
\begin{center}
\mbox{
(a)
\hspace*{-1.cm}\includegraphics[height=.3\textheight, angle =0]{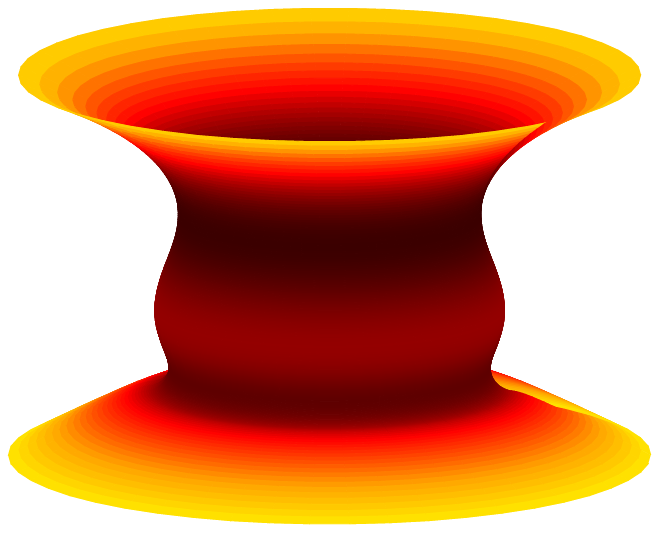}
(b)
\hspace*{-3.cm}\includegraphics[height=.3\textheight, angle =0]{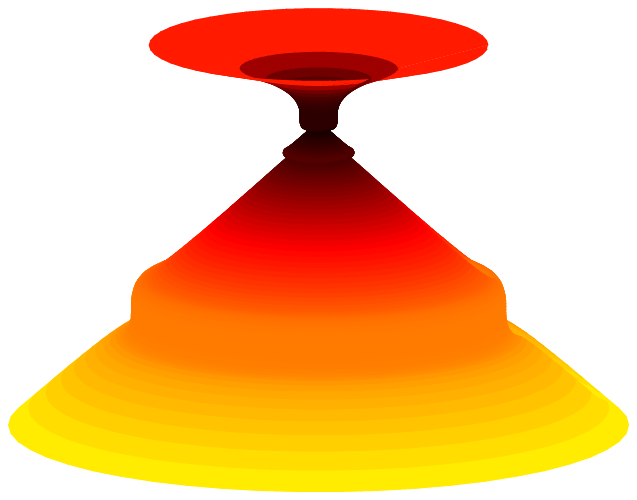}}
\end{center}
\vspace{-0.5cm}
\caption{Static wormholes in STT-1 for positive values of $\beta$:
Isometric embedding of the equatorial plane %in the Einstein frame (a) and
in the Jordan frame for
$\beta=1.0$, $Q_\psi/\eta_0=2.5$, $M_E/\eta_0=0.75$, $a=0.5$ (a)
%: embEthqrIa.eps (Fig5a.eps)
and
$\beta=2.5$, $Q_\psi/\eta_0=4.0$, $M_E/\eta_0=1.732$, $a=0.5$ (b).
%: embEthqrIb.eps (Fig5b.eps)
\label{Fig5}
}
\end{figure}

Let us now turn to the case of positive $\beta$. 
In Fig.\ref{Fig4} we illustrate the equatorial radius of the throat
in the Jordan frame $\tilde R_e$ (top) and the scalar fields (bottom) 
for two sets of static scalarized wormhole solutions with varying $\beta$. 
Clearly, the oscillating behavior of the gravitational scalar field 
translates into an oscillating behavior or the equatorial radius
of the throat in the Jordan frame.
The larger the value of $\beta$ is chosen, the more
oscillations manifest in $\tilde R_e$.
Thus the presence of a single throat (and no equator) in the Einstein frame
can lead to wormholes with multiple throats and equators in the Jordan frame.
We show in Fig.\ref{Fig5} an isometric embedding of the equatorial plane
of two wormhole solutions in the Jordan frame,
highlighting the oscillating behavior or the equatorial radius.

%In contrast to the case of 
%negative $\beta$, the wormholes in the Jordan frame can possess multiple
%throats and equators, as shown in Fig.~\ref{Fig4} for some examples.
%Fig.~\ref{Fig5} 

\subsection{Model STT-2}

As our second example we choose for the non-minimal coupling function
${\cal A} = e^{\alpha\varphi}$.
Thus the model corresponds to Brans-Dicke theory \cite{Brans:1961sx},
for which there are rather stringent observational bounds for its parameter $\alpha$,
$\alpha< 4 \cdot 10^{-3}$
\cite{Freire:2012mg}.

\subsubsection{Solutions}

From Eq.~(\ref{redequphi}) we find in this case for the gravitational scalar field the second order equation
\begin{equation}
\partial_\eta(q \partial_\eta \varphi) =
-\alpha e^{-2\alpha\varphi}\frac{Q_\psi^2}{q} 
\Longleftrightarrow
\partial_{x}^2 \varphi =-\alpha e^{-2\alpha\varphi}\left(\frac{Q_\psi}{\eta_0}\right)^2
\label{redequphi1d}
\end{equation}
Thus the rhs does not vanish for $\varphi=0$.
Hence the solutions of General Relativity are not solutions of STT-2.

To find wormhole solutions of STT-2 
we consider the first order equation 
\begin{equation}
\partial_{x} \varphi 
= \frac{Q_\psi}{\eta_0} e^{-\alpha\varphi}
\sqrt{1-a^2 e^{2\alpha\varphi} }  
\ ,
\label{equphi2x2}
\end{equation}
which is consistent with Eq.~(\ref{redequphi1d}).\\
This ODE has solutions in closed form
%
%\begin{equation}
%\varphi(x)= \frac{1}{\alpha}{\rm ln}\left\{
%\cos\left(|\hat{Q}_\psi|\left\{x-\frac{\pi}{2}\right\}\right)
%+\frac{\sqrt{1-a^2}}{a}\sin\left(|\hat{Q}_\psi|\left\{x-\frac{\pi}{2}\right\}\right)
%\right\} \ ,
%\label{phisold}
%\end{equation}
%with $\hat{Q}_\psi = \alpha a Q_\psi/\eta_0$, which satisfy the condition 
%$\varphi(x=\pi/2)=0$
%
\begin{equation}
\varphi(x)= \frac{1}{\alpha}{\rm ln}\left[
\frac{1}{\cs}\cos\left(\cs\hat{Q}_\psi\left\{\frac{\pi}{2}-x\right\}+\sigma\right)
\right] \ 
\label{phisold}
\end{equation}
with $\hat{Q}_\psi = \alpha Q_\psi/\eta_0$ and $\cs =a$, which satisfy the condition 
$\varphi(x=\pi/2)=0$. 

For the phantom scalar field we find the first order ODE
\begin{equation}
\partial_x \psi = 
\frac{\hat{Q}_\psi}{\alpha}
\frac{\cos^2\sigma}
{
\cos^2\left(\cs\hat{Q}_\psi\left\{\frac{\pi}{2}-x\right\}+\sigma\right)
} \ .
\label{psieqx2}
\end{equation}
Integration yields 
\begin{equation}
\psi(x) = 
-\frac{\cs}{\alpha}\left[
\tan\left(\cs\hat{Q}_\psi\left\{\frac{\pi}{2}-x\right\}+\sigma\right)
-\tan\sigma \right] \ ,
\label{psisolx2}
\end{equation}
where the integration constant has been chosen such that 
$\psi(x=\pi/2)=0$.

Let us now consider the solutions in the Jordan frame. 
With $F(\Phi) = {\cal A}^{-2} = e^{-2\alpha\varphi}$
and $d {\rm ln} F/d\Phi = -2\alpha d\varphi/d\Phi$
we find from Eq.~(\ref{CONF2})
\begin{equation}
\left(\frac{d\left(e^{-\alpha \varphi}\right)}{d\Phi}\right)^2
=\frac{1}{2}\frac{\alpha^2}{1-3 \alpha^2} Z(\Phi)
\label{CONF2di} \ .
\end{equation}
This implies $Z(\Phi)>0$ if $\alpha^2<3$, 
and $Z(\Phi)<0$ if $\alpha^2>3$,
respectively.
In the simple case of constant $Z(\Phi)$, $|Z(\Phi)|=Z_0^2$, this yields
\begin{eqnarray}
\Phi_\pm(x) & = & \pm \frac{\sqrt{2|1-3\alpha^2|}}{\alpha Z_0}
\left(e^{-\alpha\varphi(x)}-1\right)
\label{CONF2dia}\\
& = & \Phi_{\pm \, 0}
\left(
\frac{\cs}{\cos\left(\cs\hat{Q}_\psi\left\{\frac{\pi}{2}-x\right\}+\sigma\right)}-1
\right) \ , 
\label{solPhid} \\
F(\Phi) & = & \left(1+\frac{\Phi_\pm}{\Phi_{\pm \, 0}}\right)^2 \ ,
\label{solFd}
\end{eqnarray}
with $\Phi_{\pm \, 0}=\pm\frac{\sqrt{2|1-3\alpha^2|}}{\alpha Z_0}$.

The scalar charge $Q_\Phi$ can be computed from
\begin{equation}
Q_\Phi = \eta_0 \left. \partial_x \Phi\right|_{x=\pi/2}
 = -\alpha \Phi_{\pm \, 0} \eta_0 \left. \partial_x \varphi\right|_{x=\pi/2}
 = -\alpha \Phi_{\pm \, 0} Q_\varphi \ .
 \label{CharPhid} 
\end{equation}

\subsubsection{Domain of existence}

%The domain of existence is determined from
%$|\hat{Q}_\psi| \geq |\hat{Q}_\phi|$ and from the condition
%$e^{\alpha\varphi}>0$. Since $\varphi(x)$ is a monotonic function,
%this leads to 
%%
%\begin{eqnarray}
%-\frac{1+\frac{2}{\pi}\arccos(a)}{2a} < \hat{Q}_\psi < \frac{1-\frac{2}{\pi}\arccos(a)}{2a}
%& & {\rm for} \ a > 0 \ ,
%\label{psi_doe1}\\
%\frac{3-\frac{2}{\pi}\arccos(a)}{2a} < \hat{Q}_\psi < \frac{1-\frac{2}{\pi}\arccos(a)}{2a}
%& & {\rm for} \ a < 0 \ .
%\label{psi_doe2}
%\end{eqnarray}
%
%The domain of existence in the $\hat{Q}_\psi-\hat{Q}_\varphi$ plane is shown
%in Fig.\ref{Fig6}. 
%Note, that $\hat{Q}^{\rm cr}_\varphi \to \pm\hat{Q}^{\rm cr}_\psi$
%in the limit $\hat{Q}^{\rm cr}_\psi \to -\infty$.
The domain of existence is determined from
$|\hat{Q}_\psi| \geq |\hat{Q}_\phi|$ and from the condition
$e^{\alpha\varphi}>0$.
For convenience we write the solution
as
\begin{equation}
e^{\alpha \varphi} = \chi 
= \frac{\cos\left(y\alpha D/\eta_0+\sigma\right)}{\cos\sigma} \ ,
\end{equation}
where $y=\pi/2 -x$.
We note that $\chi$ does not change when we add any integer times $\pi$ 
to $\sigma$. Therefore it is sufficient to restrict to
$0\leq \sigma \leq \pi$. However, we have  to exclude $\sigma =\pi/2$.
The domain of existence is determined by the condition 
$\chi(y) > 0$ for $0\leq y \leq \pi$.
First we note that this condition implies $|\alpha D/\eta_0|\leq 1$,
otherwise $y\alpha D/\eta_0$ would cover an interval larger than $\pi$ and 
$\chi$ would vanish at some point, no matter what value of $\sigma$ is
chosen.

\begin{figure}[t!]
\begin{center}
\mbox{
(a)\includegraphics[height=.23\textheight, angle =0]{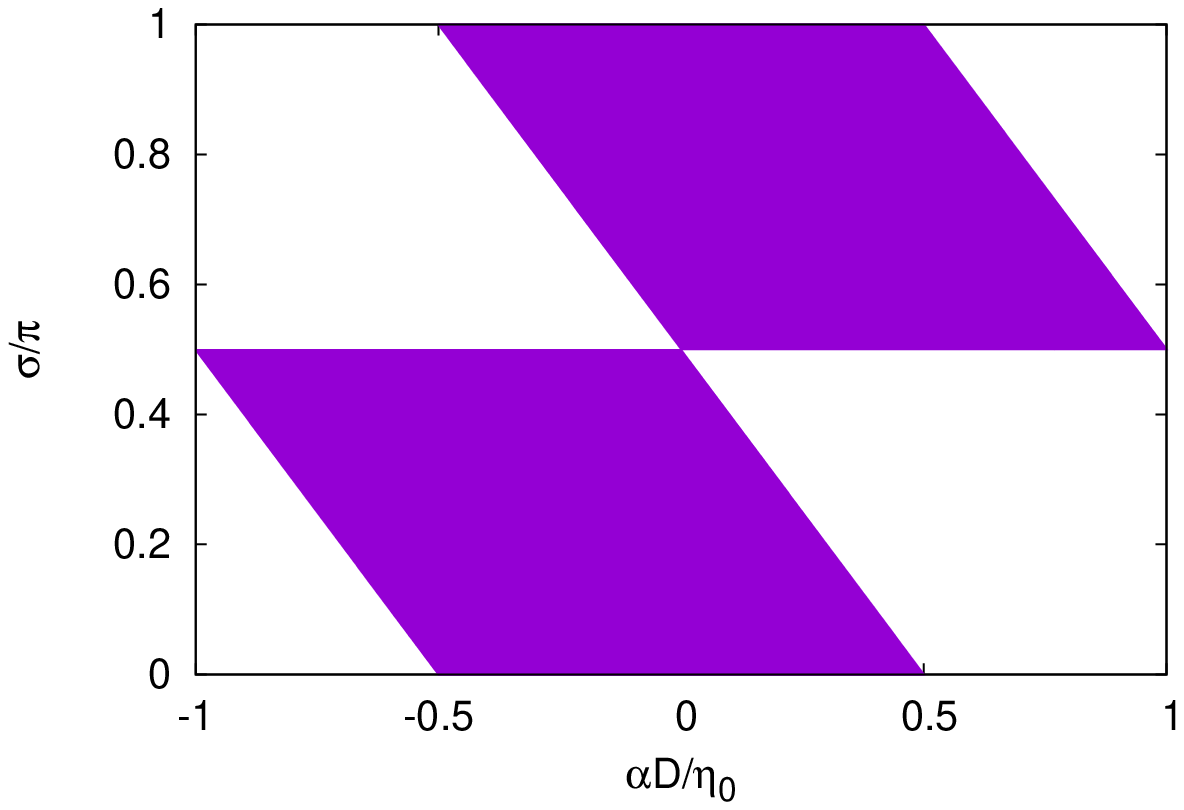}
(b)\includegraphics[height=.23\textheight, angle =0]{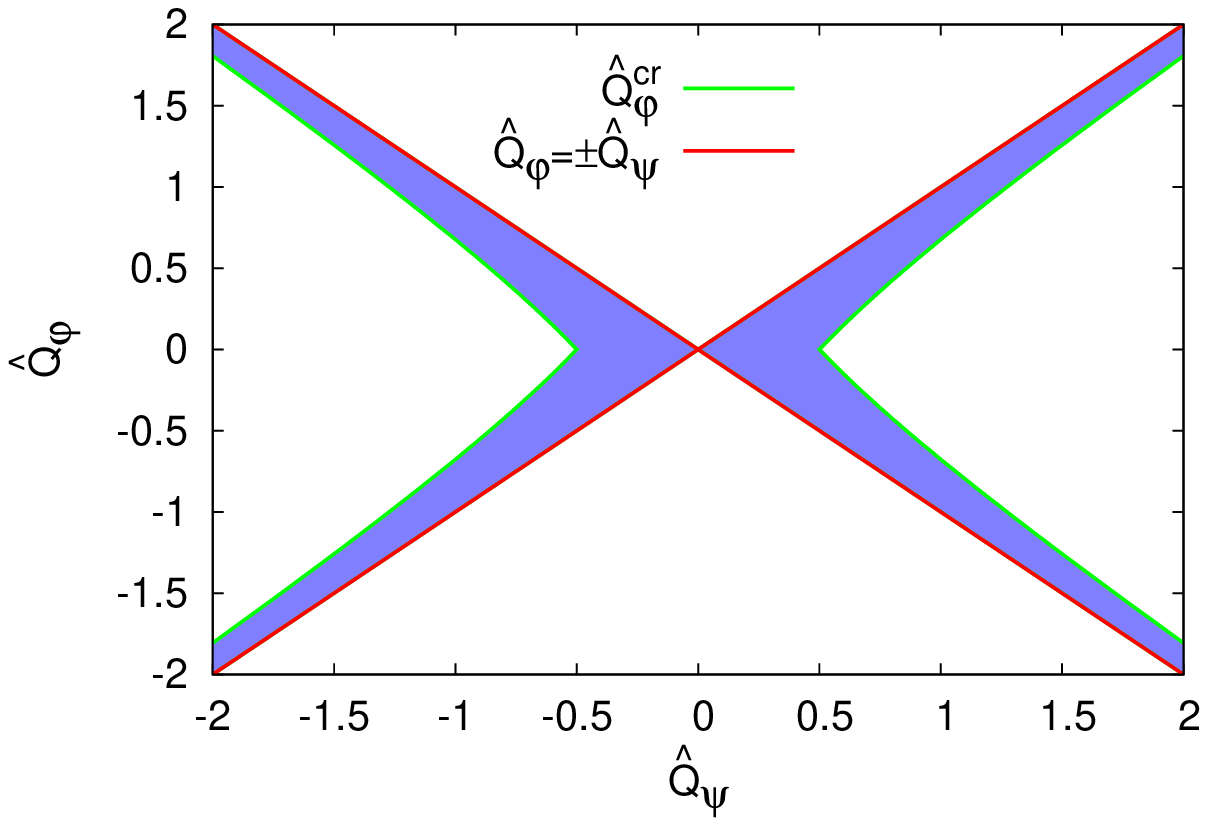}
}
\end{center}
\vspace{-0.5cm}
\caption{Wormholes in STT-2:
The domain of existence (blue)
in the $\alpha D/\eta_0-\sigma$ plane (a)
in the $\hat{Q}_{\psi}$-$\hat{Q}_{\varphi}$ plane
with $\hat{Q}_{\psi}$ again the abscissa (b).
The lines indicate the limits $\hat{Q}_{\varphi} = \pm \hat{Q}_{\psi}$ (red)
and $\hat{Q}_{\varphi}^{\rm cr}$ (green).
\label{Fig6}
}
\end{figure}

Let us now turn to the limits of $\sigma$.
We start with $0\leq \sigma < \pi/2$, when $\cos\sigma > 0$.
In this case the domain of existence is determined by
\begin{equation}
-\frac{\pi}{2}  <  y\frac{\alpha D}{\eta_0} +\sigma   <\frac{\pi}{2} \ , \ \ \ 
{\rm for} \ 0 \leq y \leq \pi \ .
\end{equation}
For $y=0$ this condition is already satisfied. For 
$y=\pi$ this yields 
\begin{equation}
-\frac{\pi}{2} -\pi\frac{\alpha D}{\eta_0} < \sigma   <\frac{\pi}{2} -\alpha D \pi \ .
\end{equation}
Employing the condition $0\leq \sigma < \pi/2$
we find 
\begin{equation}
{\rm max}\left(-\frac{\pi}{2} -\pi\frac{\alpha D}{\eta_0}, 0\right)
< \sigma   <
{\rm min}\left(\frac{\pi}{2} -\pi\frac{\alpha D}{\eta_0}, \frac{\pi}{2}\right) \ .
\end{equation}
Similarly we find for the case  $\pi/2 < \sigma \leq \pi$
\begin{equation}
{\rm max}\left(\frac{\pi}{2} -\pi\frac{\alpha D}{\eta_0} , \frac{\pi}{2}\right)
< \sigma   <
{\rm min}\left(\frac{3\pi}{2} -\pi\frac{\alpha D}{\eta_0}, \pi\right) \ .
\end{equation}
The domain of existence in the $\alpha D/\eta_0-\sigma$ plane is shown in 
Fig.~\ref{Fig6}(a)
by the blue areas.
In order to find the domain of existence for the charges we express
the scaled quantities $\hat{Q}_\psi$ and 
$\hat{Q}_\varphi$ in terms of $\alpha D/\eta_0$ and $\sigma$
\begin{equation}
\hat{Q}_\psi=\alpha \frac{Q_\psi}{\eta_0} = \pm\alpha \frac{D}{\eta_0\cos\sigma} \ , \ \ \ 
\hat{Q}_\varphi=\alpha \frac{Q_\varphi}{\eta_0} = \pm\alpha \frac{D}{\eta_0 \tan \sigma} \ .
\end{equation}
The domain of existence in the $\hat{Q}_\psi-\hat{Q}_\varphi$ plane is 
shown in Fig.~\ref{Fig6}(b).

For the static wormholes the charges and the mass are related by
\begin{equation}
Q_\psi^2-Q_\varphi^2 = \eta_0^2 + M_{\rm E}^2 \ .
\label{chargemassstat}
\end{equation}
This leads to an additional reduction of the domain of existence,
\begin{equation}
Q_\psi^2-Q_\varphi^2 \geq  \eta_0^2  \ .
\label{DoEstat}
\end{equation}

\subsubsection{Properties}

Next we consider the mass, the angular momentum and the quadrupole moment
in STT-2. 
As discusssed in Section {2.5} the angular momentum 
is the same in both frames.
In the asymptotic region $\Sigma_+$ the coupling function ${\cal A}$ 
satisfies $d{\rm ln}{\cal A}/d\varphi = \alpha\neq 0$, 
thus the gravitational mass and the Schwarzschild
mass in the Jordan frame differ from the ADM mass in the Einstein frame,
\begin{equation}
M_{\rm K+} - M_{\rm S+} = 2\alpha Q_\varphi  \ .
\label{d-STTmass}
\end{equation}
In STT-2 the quadrupole moments differ in the Jordan frame and in the
Einstein frame,
\begin{equation}
{\cal Q}_{\rm J+} = {\cal Q}_{\rm E+} +\frac{1}{3}\alpha Q_\varphi c_2 \ .
\label{d-STTquad}
\end{equation}

In the asymptotic region $\Sigma_-$ the gravitational scalar field assumes 
a finite value $\varphi_-$. The conformal factor and the scalar charges 
in $\Sigma_-$ can be expressed in terms of $\hat{Q}_\psi$ and $a$ as
\begin{eqnarray}
{\cal A}_-^2  & = & \frac{\cos^2\left(\cs\pi\hat{Q}_\psi+\sigma\right)}{\cos^2\sigma} \ ,
\label{d-STTAmin}\\
Q_{\psi -} & = & Q_{\psi}\frac{\cos^2\sigma}{\cos^2\left(\cs\pi\hat{Q}_\psi+\sigma\right)} \ ,
\label{d-STTQpsimin}\\
Q_{\varphi -} & = & Q_{\psi}\cs  \tan\left(\cs\pi\hat{Q}_\psi+\sigma\right) \ .
\label{d-STTQphimin}
\end{eqnarray}
This yields for the masses in the Jordan frame 
\begin{eqnarray}
M_{\rm K-} + M_{\rm S-} & = &
\frac{2}{\cs}\cos\left(\cs\pi\hat{Q}_\psi+\sigma\right)
\bar{M}_{\rm E-} \ ,
\label{d-STTmassmin1}
\\
M_{\rm K-} - M_{\rm S-}  & = &
-2\alpha {Q}_\psi e^{-\frac{\gamma}{2}}
\sin\left(\cs\pi\hat{Q}_\psi+\sigma\right) \ ,
\label{d-STTmassmin2}
\end{eqnarray}
where $\bar{M}_{\rm E-}$ denotes the mass in the Einstein frame.

The angular momentum in the Jordan frame differs from the angular momentum in
the Einstein frame by some factor,
\begin{equation}
J_{\rm J-}=\bar{J}_{\rm E-} \frac{1}{\cos^2\sigma}\cos^2\left(\cs\pi\hat{Q}_\psi+\sigma\right).
\label{d-STTangmin}
\end{equation}
For the quadrupole moment we find from the general expression, 
Eq.~(\ref{Quadrelationmin}),
\begin{equation}
{\cal Q}_{\rm J-} =  \frac{1}{\cos^3\sigma}\cos^3\left(\cs\pi\hat{Q}_\psi+\sigma\right)
\left(
{\cal Q}_{E-}
+\frac{\alpha}{3} Q_{\varphi-} e^{-\frac{\gamma}{2}}\bar{c}_2
\right) \ .
\label{d-STTquadmin}
\end{equation}

Considering the geometric properties 
of the wormholes in the Jordan frame,
%In STT-2 
the circumferential radius 
and the condition for the throat coordinate read
\begin{equation}
\tilde{R}_e(x) = 
\frac{\eta_0}{\cos x} \left. e^{\alpha \varphi-f/2} \right|_{\theta=\pi/2} \ , \ \ \ \
\left(\tan x +\alpha\partial_x \varphi-\frac{1}{2}\partial_x f
\right)_{x_t, \theta=\pi/2} =0 \ .
\label{d-STTthroat}
\end{equation}

Let us remark, that the static wormhole solutions of Brans-Dicke
theory obtained without a phantom field 
%are not traversable in practice, and 
do not exist in the Einstein frame
\cite{Agnese:1995kd,Nandi:1997mx,Nandi:1997en}.

\subsection{Model STT-3}

As our third example we consider the coupling function
${\cal A}(\varphi)=\cosh(\varphi/\sqrt{3})$.
We are not aware of any previous investigations with this coupling function.

\subsubsection{Solutions}

From Eq.~(\ref{redequphi}) we find for the gravitational scalar field the equation
\begin{equation}
\partial_\eta \left(\hq \partial_\eta \varphi\right) =
-\frac{1}{\sqrt{3}}
\frac{\sinh\left(\frac{\varphi}{\sqrt{3}}\right)}{\cosh^3\left(\frac{\varphi}{\sqrt{3}}\right)}
 \frac{1}{\hq}Q_\psi^2 
\Longleftrightarrow
\partial_x^2 \varphi =
-\frac{1}{\sqrt{3}}
\frac{\sinh\left(\frac{\varphi}{\sqrt{3}}\right)}{\cosh^3\left(\frac{\varphi}{\sqrt{3}}\right)}
\left(\frac{Q_\psi}{\eta_0}\right)^2 
 \ ,
\label{redequphiex3}
\end{equation}
which allows for the trivial solution $\varphi =0 $, 
present in General Relativity.
In order to obtain non-trivial solutions 
for the scalar field we turn to the first order equation,
\begin{equation}
\partial_{x} \varphi 
= \frac{Q_\psi}{\eta_0} 
\frac
{\sqrt{1-a^2 \cosh^2(\frac{\varphi}{\sqrt{3}})}}
{\cosh\left(\frac{\varphi}{\sqrt{3}}\right) } \ ,
\label{equphiex3}
\end{equation}
which is consistent with Eq.~(\ref{redequphiex3}).\\
This ODE has solutions 
\begin{equation}
\varphi(x)=
\sqrt{3}{\rm arsinh}\left[\frac{\sqrt{1-a^2}}{a}
\sin\left(\frac{a}{\sqrt{3}}\frac{Q_\psi}{\eta_0}\left\{x-\frac{\pi}{2}\right\}\right)
\right] \ ,
\label{solphiex3}
\end{equation}
which obey the condition $\varphi(\pi/2)=0$.

The ODE of the phantom field becomes
\begin{equation}
\partial_x \psi = \frac{Q_\psi}{\eta_0}
\frac{1}
{
1+\frac{1-a^2}{a^2}
\sin^2\left(\frac{a}{\sqrt{3}}\frac{Q_\psi}{\eta_0}\left\{x-\frac{\pi}{2}\right\}\right)
} \ ,
\label{eqpsiex3}
\end{equation}
and has the solution
\begin{equation}
\psi(x) = 
\sqrt{3}\arctan\left[\frac{1}{a}
\tan\left(\frac{a}{\sqrt{3}}\frac{Q_\psi}{\eta_0}\left\{x-\frac{\pi}{2}\right\}\right)
\right] \ ,
\label{solpsiex3}
\end{equation}
which satisfies the boundary condition $\psi(\pi/2)=0$.

Let us now go to the Jordan frame. With $F(\Phi)={\cal A}(\varphi)^{-2}$ and $Z(\Phi)=1$
we find from 	
Eq.~(\ref{CONF2})
\begin{eqnarray}
F(\Phi) & = & 1-\frac{\Phi^2}{6}  \ ,
\label{a3}\\
\Phi(\varphi) & = & \pm \sqrt{6}\tanh\left(\frac{\varphi}{\sqrt{3}}\right) \ .
\label{Phi3}
\end{eqnarray}

Substitution of the solution Eq.~(\ref{solphiex3}) in Eq.~(\ref{Phi3}) gives the gravitational
scalar field in the Jordan frame
\begin{equation}
\Phi(x)= \sqrt{6}\frac{
\sqrt{1-a^2}
\sin\left(\frac{a}{\sqrt{3}}\frac{Q_\psi}{\eta_0}\left\{x-\frac{\pi}{2}\right\}\right)
}
{\sqrt{
a^2+(1-a^2)
\sin^2\left(\frac{a}{\sqrt{3}}\frac{Q_\psi}{\eta_0}\left\{x-\frac{\pi}{2}\right\}\right)}
} \ .
\label{Phi3a}
\end{equation}
For the scalar charge in the Jordan frame we find 
$Q_\Phi = \sqrt{2} Q_\varphi = \sqrt{2} \sqrt{1-a^2} Q_\psi$.
Moreover, substitution in $F(\Phi) = 1-\Phi^2/6$ yields
\begin{equation}
F(x) = \frac{a^2}
{
a^2 \cos^2\left(\frac{a}{\sqrt{3}}\frac{Q_\psi}{\eta_0}\left\{x-\frac{\pi}{2}\right\}\right)
+\sin^2\left(\frac{a}{\sqrt{3}}\frac{Q_\psi}{\eta_0}\left\{x-\frac{\pi}{2}\right\}\right)
} \ .
\label{F3a}
\end{equation}
Thus we find that $F(\Phi) \geq 0$. We note that the solutions
are regular on the interval $-\pi/2 \leq x \leq \pi/2$ in the 
Jordan frame and in the Einstein frame as well. Thus there are no constraints
on the scalar charges except for $|Q_\psi| >|Q_\varphi|$.

\subsubsection{Properties}

\begin{figure}[t!]
\begin{center}
\mbox{
(a)
\includegraphics[height=.23\textheight, angle =0]{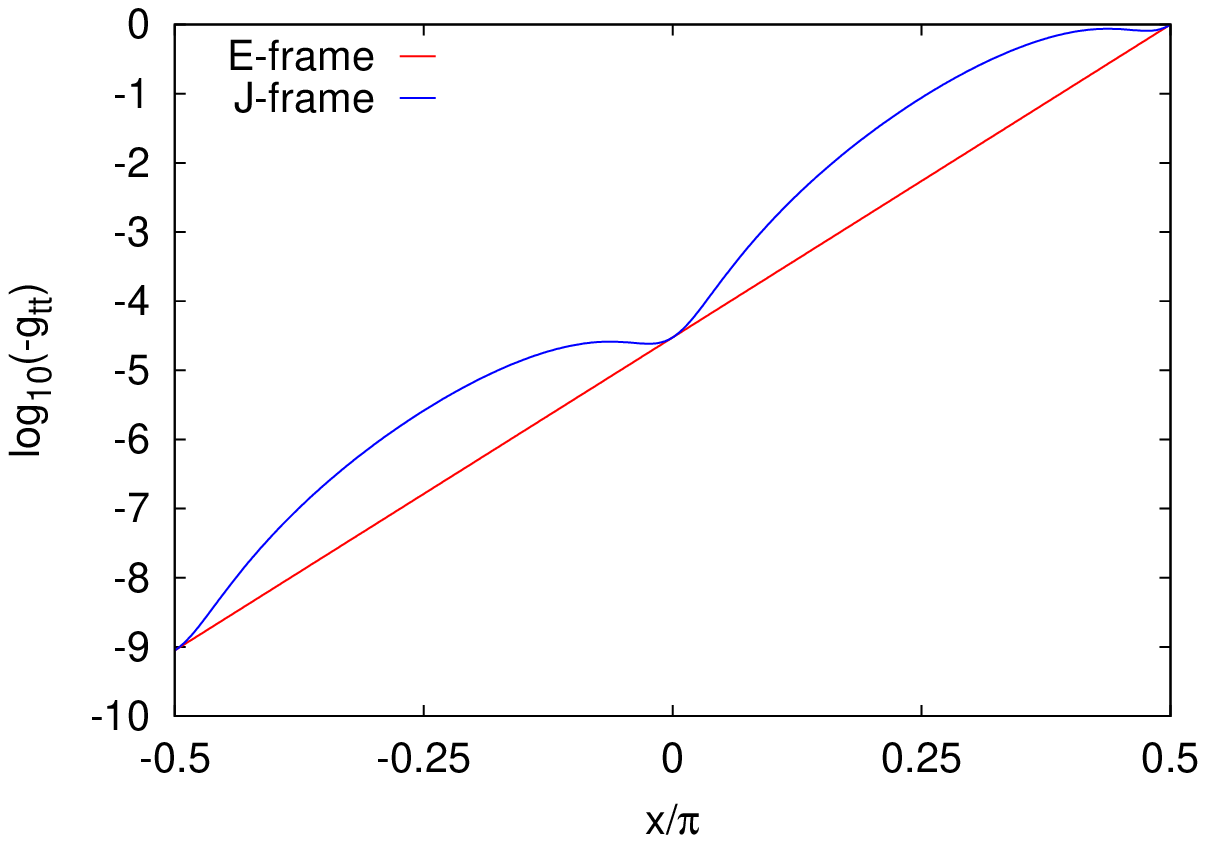}
(b)
\includegraphics[height=.23\textheight, angle =0]{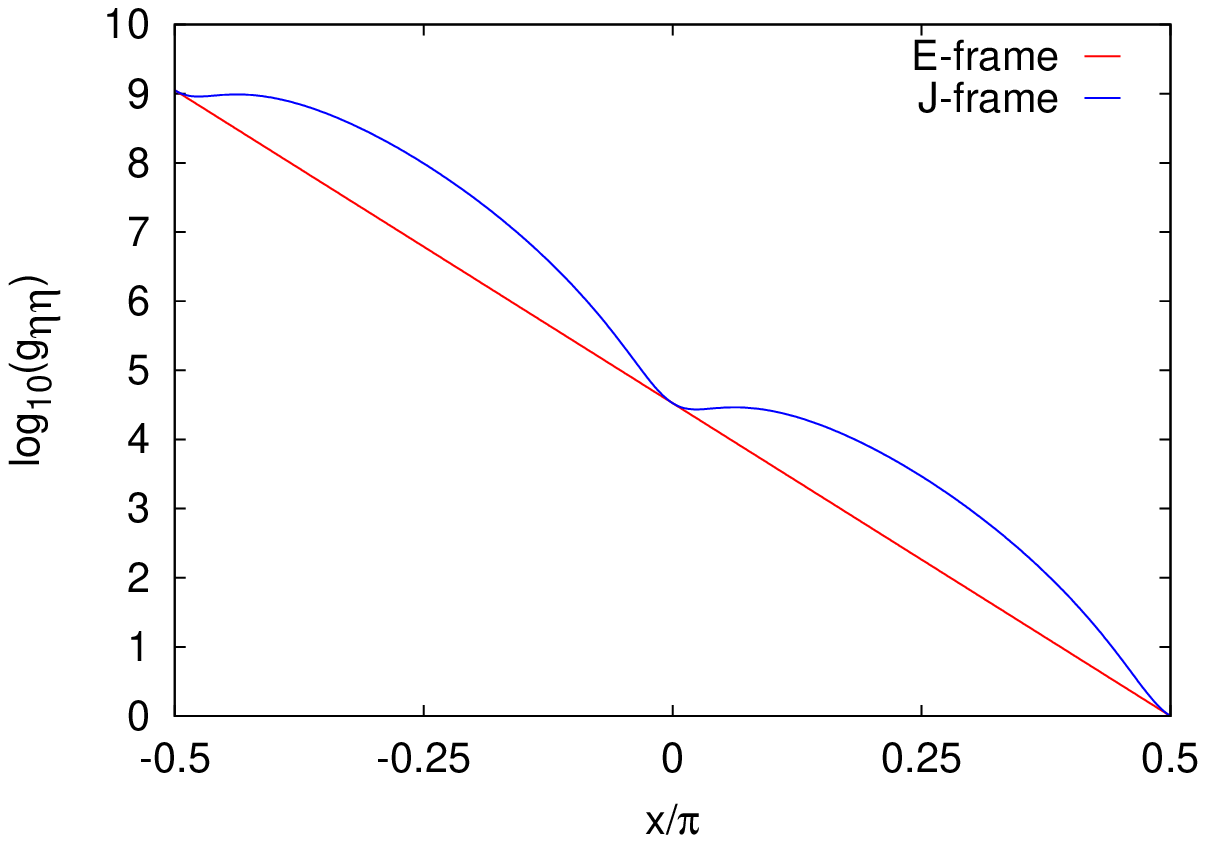}}
\mbox{
(c)
\includegraphics[height=.23\textheight, angle =0]{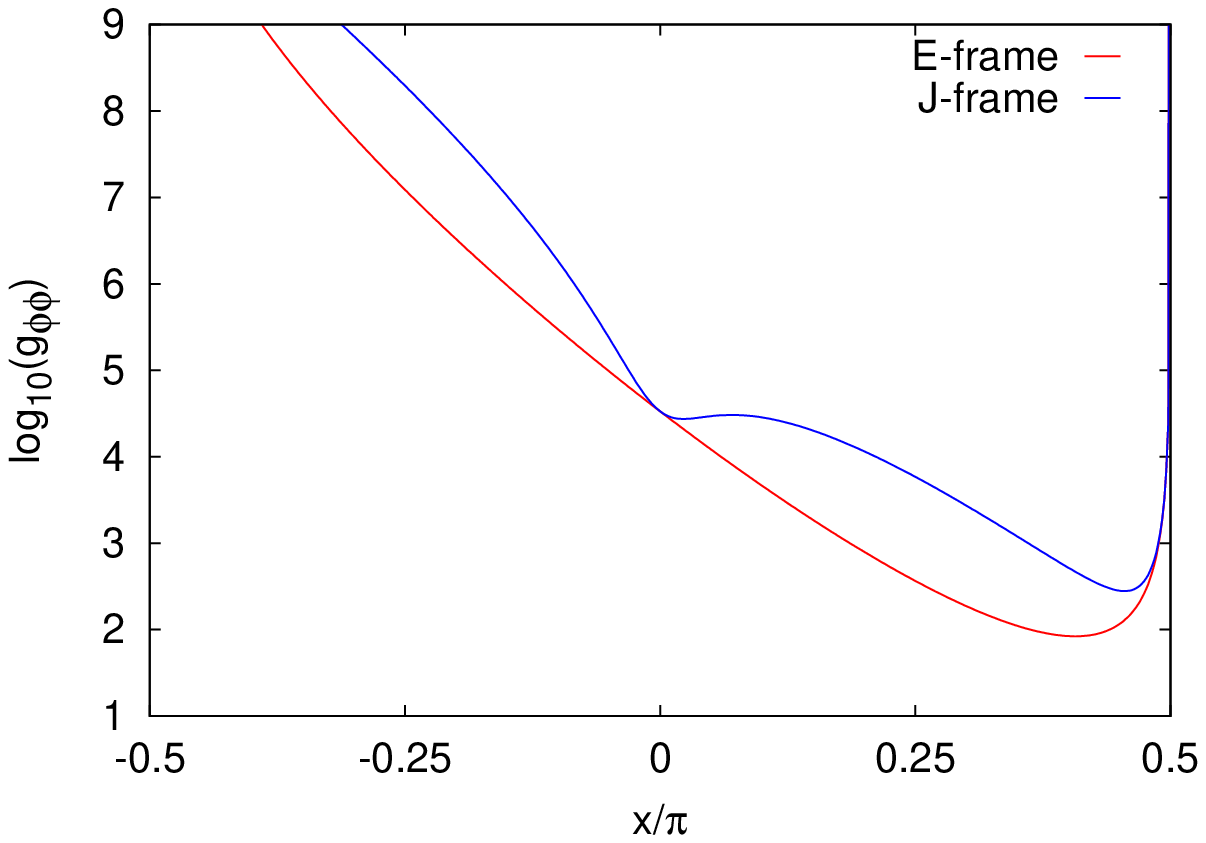}
(d)
\includegraphics[height=.23\textheight, angle =0]{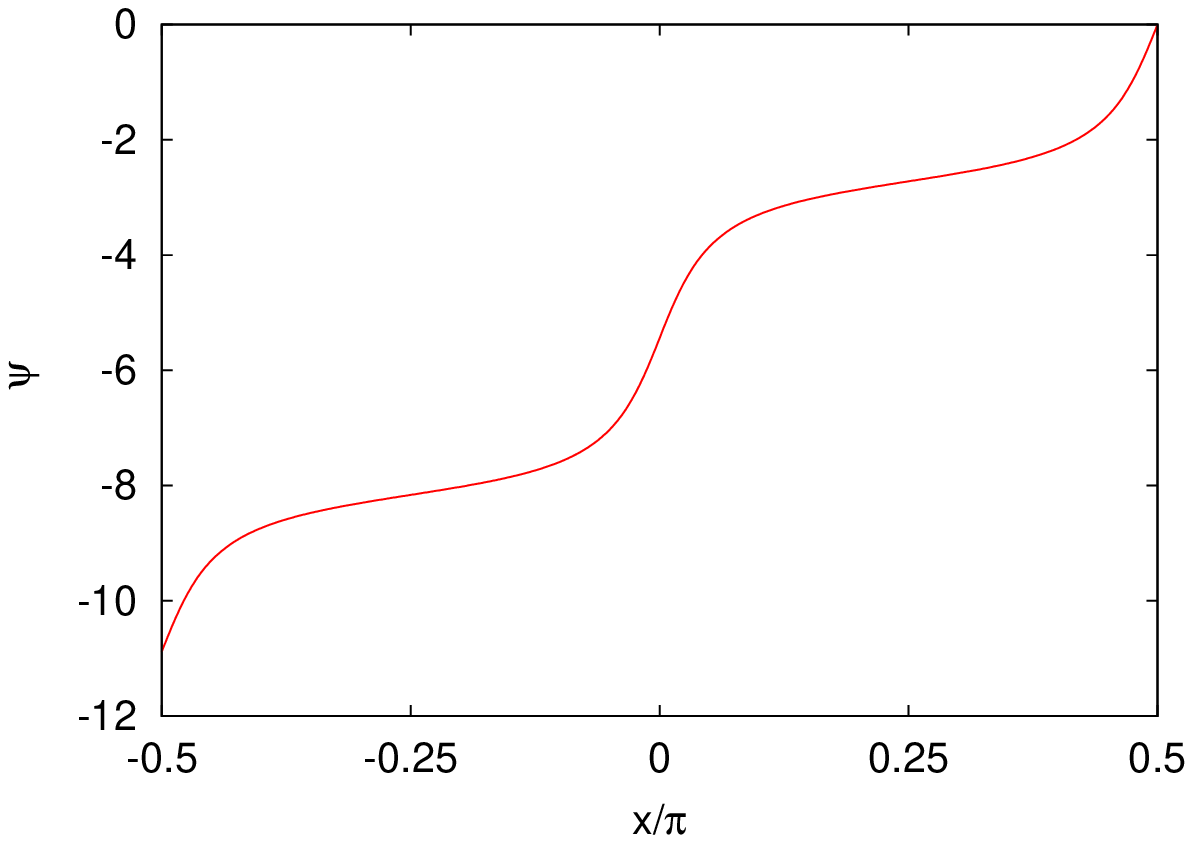}}
\mbox{
(e)
\includegraphics[height=.23\textheight, angle =0]{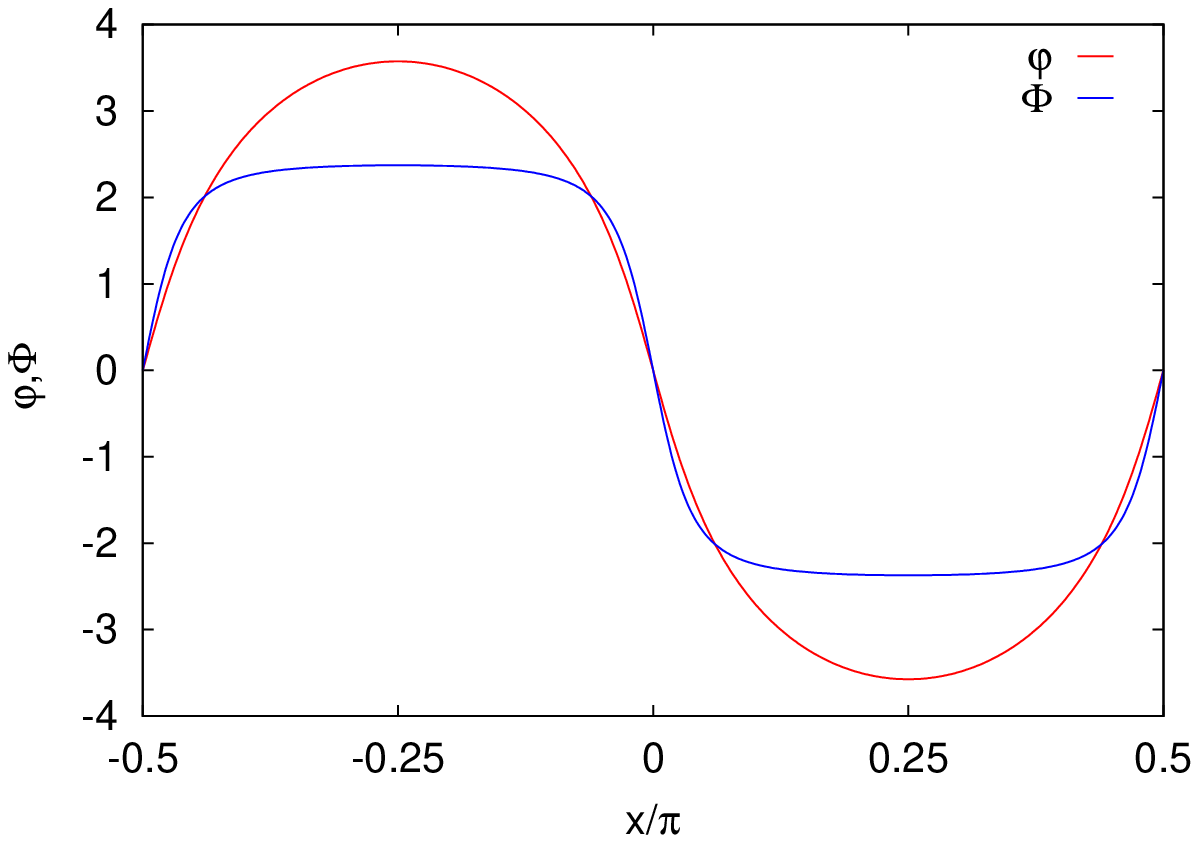}
(f)
\includegraphics[height=.23\textheight, angle =0]{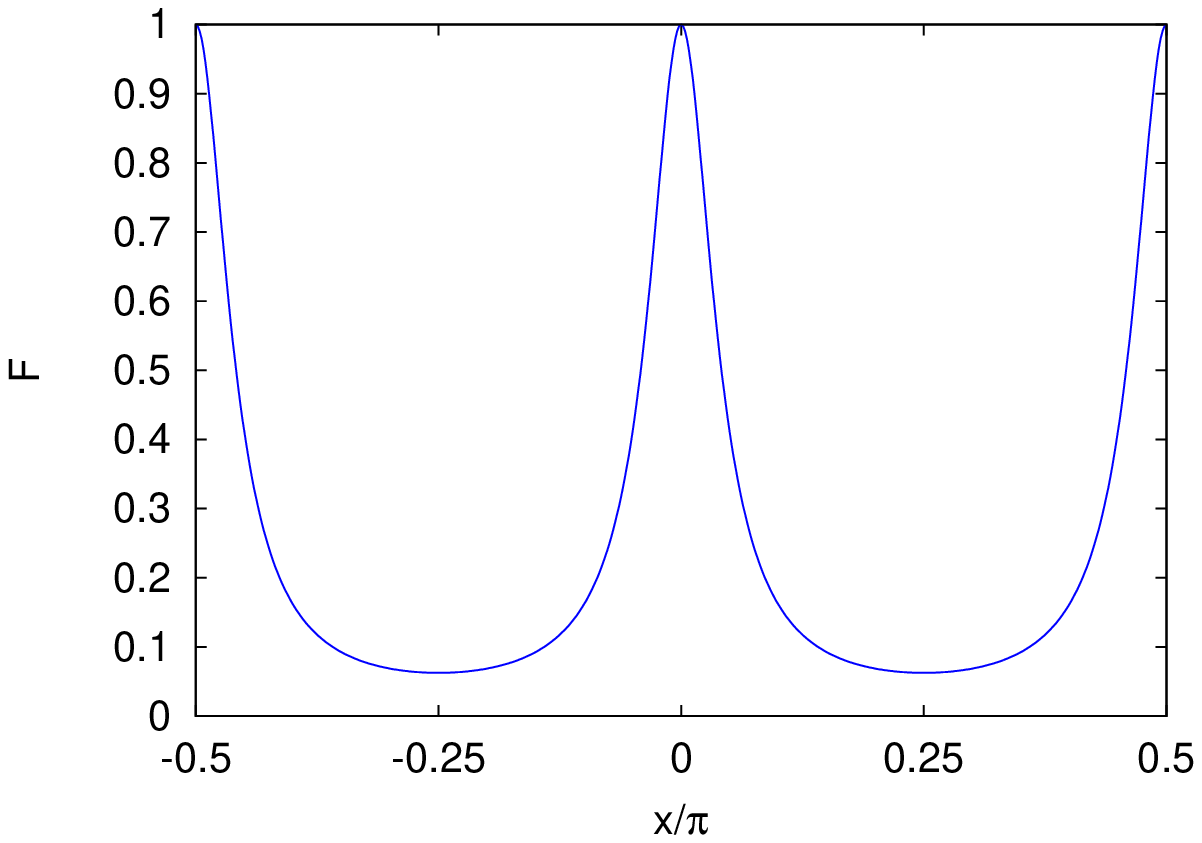}}
\end{center}
\vspace{-0.5cm}
\caption{Wormholes in STT-3:
The metric components  in the equatorial plane and the scalar fields in the Einstein frame
and in the Jordan frame, $-g_{tt}$ (a), $g_{\eta\eta}$ (b),
$g_{\phi\phi}$ (c), $\psi$ (d), $\varphi$ and $\Phi$ (e).
In (f) the function $F$ is shown.
(Parameters:
$a=1/4$, $\frac{a Q_\psi}{\sqrt{3}\eta_0}=2$.)
\label{Fig7}
}
\end{figure}

\begin{figure}[t!]
\begin{center}
\mbox{
(a)
\hspace*{-1.cm}\includegraphics[height=.3\textheight, angle =0]{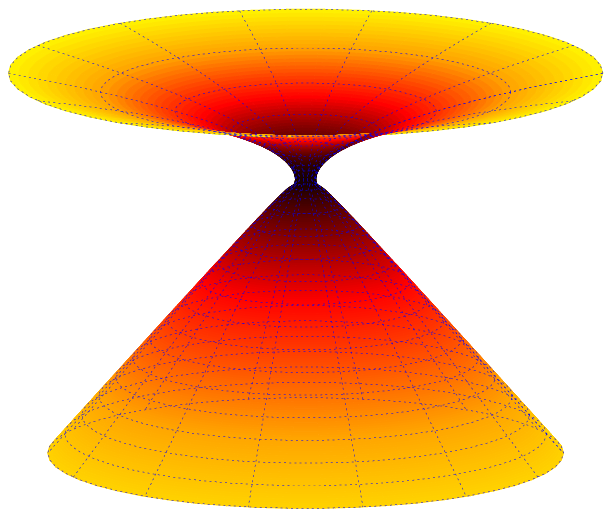}
(b)
\hspace*{-3.cm}\includegraphics[height=.3\textheight, angle =0]{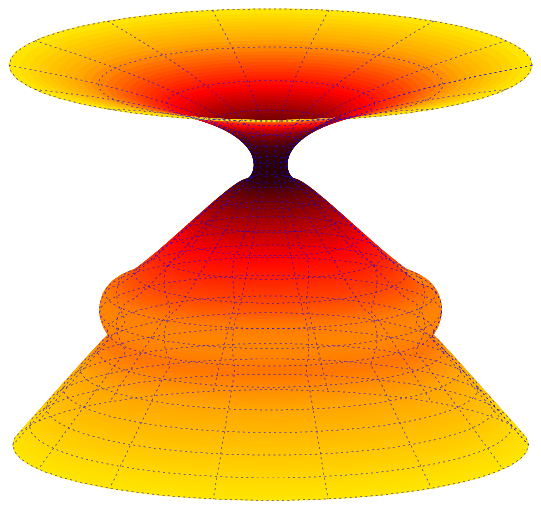}}
\end{center}
\vspace{-0.5cm}
\caption{Static wormholes in STT-3:
Isometric embedding of the equatorial plane in the Einstein frame (a)
and in the Jordan frame (b).
\label{Fig8}
}
\end{figure}

Let us now consider the geometry of the solutions.
The case $\frac{d^2}{dx^2}\tilde{R}_e(x_t) < 0$ 
corresponds to a (local) maximum of 
the circumference in the equatorial plane $\tilde{R}_e$, 
and since $\tilde{R}_e$
grows without bound in the asymptotic regions, there exist two minima
of $\tilde{R}_e$, located 
to the left and to the right of the coordinate $x_t$, respectively. 
Such a spacetime possesses two throats
with a belly (or an equator) in between them.

As an example we consider a solution which possesses a single throat in the 
Einstein frame but two throats and a belly in between in the Jordan frame.
The non-scalarized solution corresponds to the static Ellis wormhole.
The scalarized solution is given by 
Eqs.~(\ref{solphiex3}), (\ref{solpsiex3}) and (\ref{Phi3a}) and is
characterized by the parameters
$a=1/4$, $\frac{a Q_\psi}{\sqrt{3}\eta_0}=2$, 
%which is consistent with 
and $\gamma = -2\pi\sqrt{11}$, 
corresponding to $M_{E+} = -\sqrt{11}\eta_0$. 
Note, that in this example the scalar fields are anti-symmetric.
Consequently, the masses coincide in the Jordan frame and in the Einstein frame
in both asymptotically flat regions.

In Fig.\ref{Fig7} we illustrate the $tt$-, $\eta\eta$-, and $\phi\phi$- 
components of the metric (in the equatorial plane)
in the Jordan frame and in the Einstein frame. 
We observe that in the Einstein frame the $\phi\phi$-component possesses only
one minimum corresponding to a single throat. In the Jordan frame, however,
the $\phi\phi$-component possesses two mimima 
and one local maximum corresponding
to two throats and one belly, respectively.
Fig.\ref{Fig7}(d) shows the phantom field, which is the same in both frames.
The scalar fields $\varphi$ and $\Phi$ are exhibited in Fig.\ref{Fig7}(e).
Fig.\ref{Fig7}(f) shows the function $F(\Phi(x))$.

In Fig.\ref{Fig8} we show an isometric embedding of the equatorial plane
of this solution.
Here the presence of a single throat in the Einstein frame (a) and 
two throats and the belly in the Jordan frame (b) are clearly visible.

%For symmetric wormholes, the metric functions
%$f$ and $\nu$ are even functions of $\eta$.
%If there is no scalarization,
%their throat is located at the hypersurface $\eta=0$, 
%which represents a minimal surface.
%The equatorial radius $R_e$ of the throat is then given by
%\begin{equation}
%R_e = \left. \sqrt{g_{\varphi\varphi}}\right|_{\eta=0,\theta=\pi/2}
%=e^{-f_0/2} \eta_0 \ ,
%\label{Req}
%\end{equation}
%with $f_0=f(\eta=0,\theta=\pi/2)$,
%while the polar radius $R_p$ and the areal radius $R_A$ are given by
%\begin{equation}
%R_p  =\frac{\eta_0}{\pi} 
%\int_0^\pi {\left. e^{(\nu-f)/2}\right|_{\eta=0}  d\theta} \ , \ \ \
%R_A^2=\frac{\eta_0^2}{2}   
%\int_0^\pi {\left. e^{\nu/2-f}\right|_{\eta=0} \sin \theta  d\theta} \ .
%\end{equation}
%%
%Denoting the angular velocity of the throat by
%$\Omega=\omega_0$, %=\tomega(\teta=0)$,
%the rotational velocity
%of the throat in the equatorial plane is given by
%\begin{equation}
%%\frac{\Omega}{\Omega_0}=\omega_0 e^{-f_0/2} \ , \ \ \
%v_e= {R_e\Omega} \le 1 \ . % =\omega_0 e^{-f_0/2} \ , \ \ \
%%{\rm with}\ \ \Omega_0=\frac{c}{R} \ 
%\end{equation}
%%with $\Omega_0={c}/{R}$ and $\omega_0=\omega(\teta=0)$.

\begin{figure}[h!]
\begin{center}
\mbox{
(a)
\includegraphics[width=.5\textwidth, angle =0]{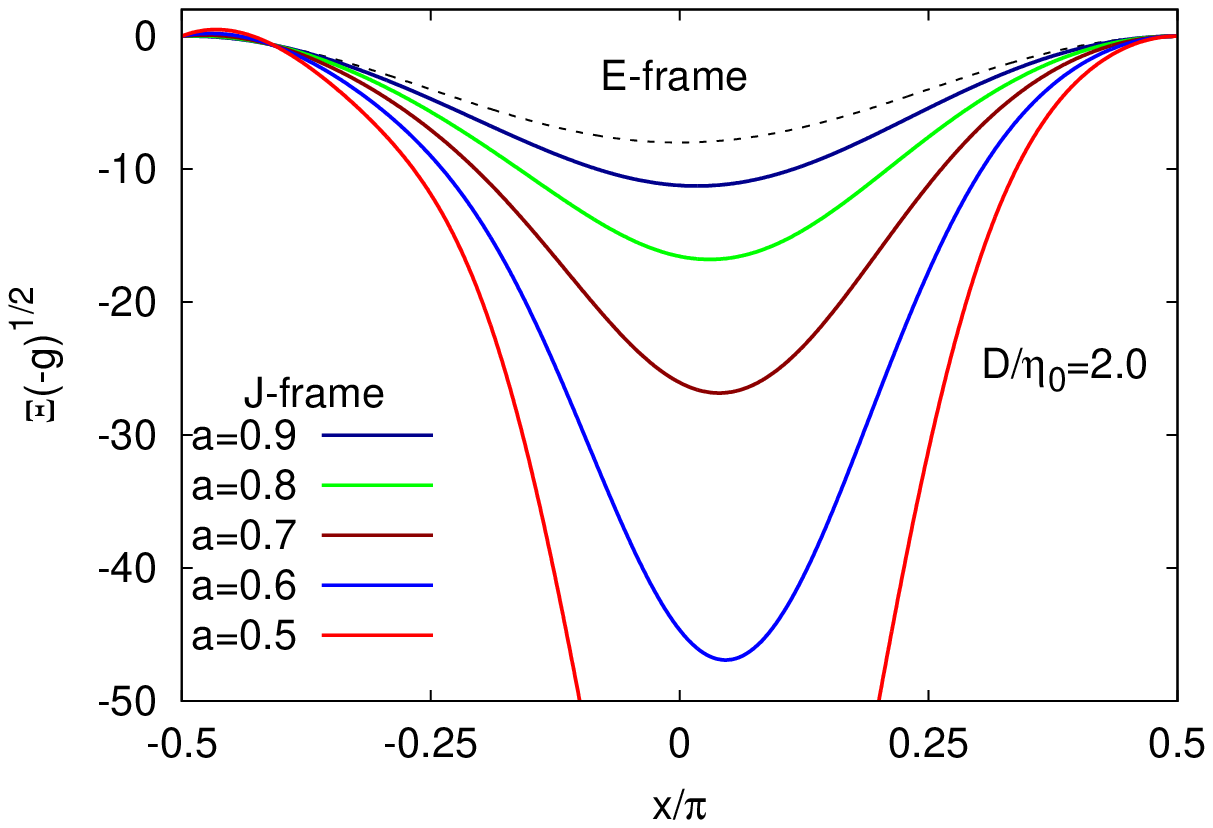}
(b)
\includegraphics[width=.5\textwidth, angle =0]{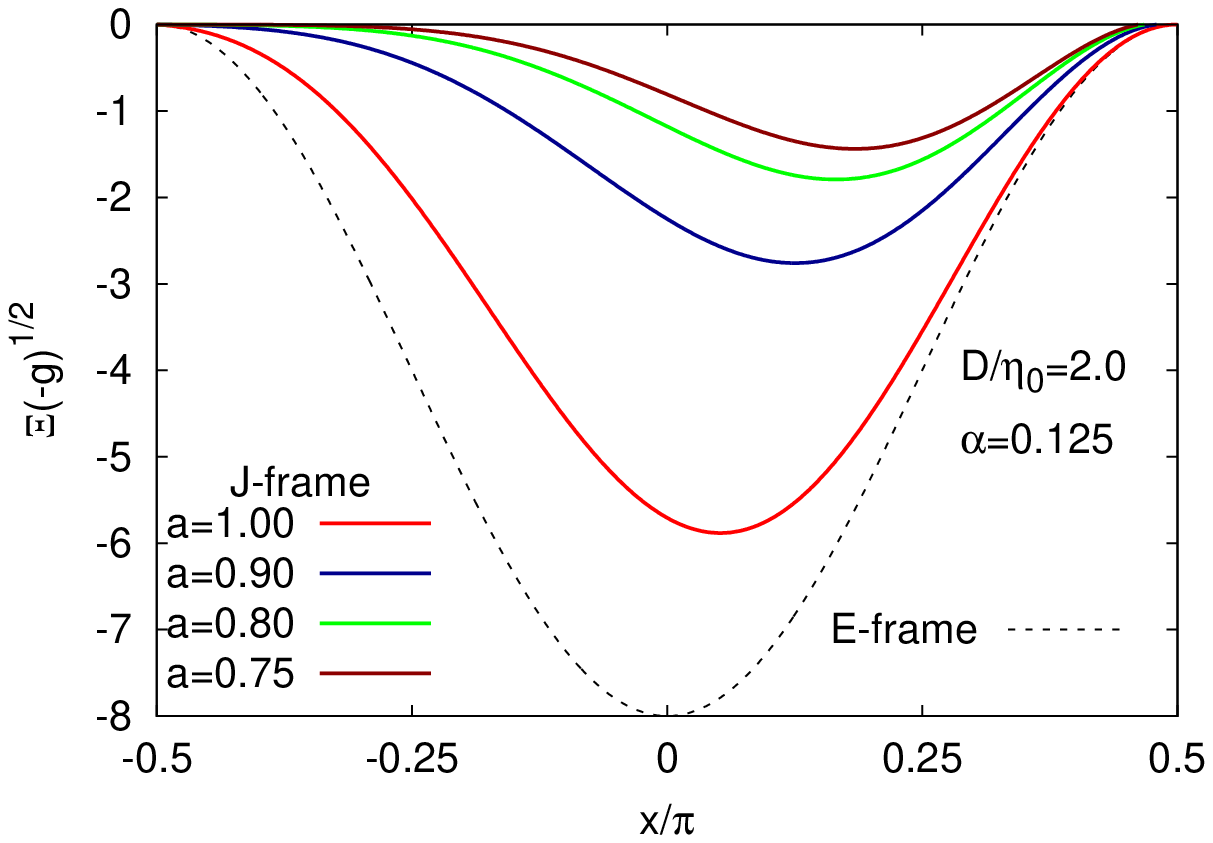}}
\end{center}
\vspace{-0.5cm}
\caption{
Violation of the NEC for static wormholes in the equatorial plane
in the Einstein and Jordan frame
for $D/\eta_0=2$ and several values of $a$ in STT-3 (a),
and in STT-2 ($\alpha=0.125$) (b).
\label{Fig9}
}
\end{figure}

In Fig.\ref{Fig9}(a) we demonstrate the violation of the NEC 
in the equatorial plane
for static STT-3 solutions for
fixed $D/\eta_0=2$ and several values of $a$.
The NEC is always violated both in the Einstein frame
and in the Jordan frame.
Clearly, the violation depends on the parameter $a$
only in the Jordan frame.
Seemingly the violation increases with decreasing $a$.
Note however, that for small values of $a$ the NEC in the Jordan frame is not violated 
in some part of the region $\Sigma_-$ extending to the asymptotic region.
For comparison we exhibit in Fig.\ref{Fig9}(b) 
the violation of the NEC
also for static STT-2 solutions ($\alpha=0.125$)
with the same value of $D/\eta_0$
and several values of $a$. 
Here we observe violation of the NEC in the full spatial domain
in both frames. But in contrast to the static STT-3 solutions
the violation of the NEC in the Jordan frame decreases with decreasing $a$.

\section{Conclusion}

In General Relativity the non-trivial topology of traversable wormholes can be
achieved by means of phantom fields, allowing for static and rotating Ellis wormholes
\cite{Ellis:1973yv,Ellis:1979bh,Bronnikov:1973fh,Kashargin:2007mm,Kashargin:2008pk,Kleihaus:2014dla,Chew:2016epf},
when no further fields are present.
In particular, the rotating Ellis wormholes possess many interesting properties, e.g.,
they satisfy a Smarr relation, they possess as limiting configuration
an extremal Kerr black hole, they possess bound orbits,  etc.

Here we have considered scalarized wormholes in STT.
We have shown, that once the 
corresponding (non-scalarized) wormhole solutions  are known in General Relativity,
it is no longer necessary to solve the Einstein equations for the metric.
But only the equations for the gravitational scalar field and the phantom scalar field
need to be solved.
Indeed, each solution in General Relativity, and thus each solution
for the metric, is characterized by a constant $D$, which relates the charges 
$Q_\varphi$ and $Q_\psi$ of the
gravitational scalar field and the phantom scalar field, respectively,  in the Einstein frame,
$D^2 = Q_\psi^2-Q_\varphi^2$.

Regarding the Jordan frame as the physical frame,
we have considered various mass definitions like the
gravitational mass, the tensor mass and the Schwarzschild mass
of the wormholes.
Assuming that the gravitational scalar field tends to zero in one asymptotically flat region,
the wormhole solutions then lead in general to a non-vanishing value of the 
gravitational scalar field in the other asymptotically flat region
(where a coordinate transformation needs to be performed to make it approach
Minkowski space).
Therefore the global charges mass and angular momentum have been
considered separately in each asymptotically flat region.

We have also derived the general expressions for the quadrupole moment
in both asymptotic regions in the Einstein frame and in the Jordan frame,
obtaining relations between the quadrupole moments in these two frames.
Likewise we have considered the geometric properties of these  wormholes
including the location of their throat(s) or equator(s) and their corresponding
circumferences. 
The NEC is always violated in the Einstein frame, independent of the
specific STT considered. In the Jordan frame the violation of the NEC
depends on the coupling function and thus on the STT. 
%In all cases
%studied, the NEC was always violated in all of space.

To give some concrete examples we have then chosen 3 specific STT,
by specifying their coupling functions ${\cal A}$: STT-1 with
${\cal A}_1(\varphi)=e^{\beta\varphi^2/2}$ \cite{Damour:1993hw,Damour:1996ke},
STT-2 with
${\cal A}_2(\varphi)=e^{\alpha\varphi}$ \cite{Brans:1961sx},
and STT-3 with
${\cal A}_3(\varphi)=\cosh(\varphi/\sqrt{3})$.
Clearly, the known Ellis wormholes of General Relativity 
are also solutions of STT-1 and STT-3, but not of STT-2.
For these STT we have solved the scalar field equations,
studied the domain of existence of the solutions
and investigated their physical properties.
Interestingly, in STT-1 scalarization arises for any positive or negative value of 
the parameter $\beta$ (recall the presence of a critical 
threshold value in the case of neutron stars and boson stars).
Here the solutions for positive values of $\beta$ may
possess multiple throats and equators in the Jordan frame,
while they possess a single throat in the Einstein frame.
Also in STT-3 we obtained wormhole solutions,
which possess a single throat in the Einstein frame,
while they possess an equator and a double throat
in the Jordan frame.

Let us end with a comment on the stability of the solutions.
Static Ellis wormholes are  known to be unstable
\cite{Shinkai:2002gv,Gonzalez:2008wd,Gonzalez:2008xk}.
Matos and Nunez  \cite{Matos:2005uh} have argued
that rotating wormholes might, however, be more stable and thus
traversable. A mode analysis of rotating wormholes in
5 dimensions has lent support to this conjecture \cite{Dzhunushaliev:2013jja}.
In 4 dimensions an analogous mode analysis 
still remains a challenge for GR wormholes,
and even more so in the case of STT wormholes.
Another road to be followed to obtain stable wormholes
could be to employ other generalized theories of gravity,
as, e.g., Einstein-Gau\ss -Bonnet-dilaton
gravity \cite{Kanti:2011jz,Kanti:2011yv}.

%\newpage

{\bf Acknowledgment}

\noindent
We gratefully acknowledge support by the DFG within the Research
Training Group 1620 ``Models of Gravity''
and by FP7, Marie Curie Actions, People, 
International Research Staff Exchange Scheme (IRSES-606096).
BK gratefully acknowledges support
from Fundamental Research in Natural Sciences
by the Ministry of Education and Science of Kazakhstan.
We gratefully acknowledge discussions with E.~Radu.

\section*{Appendix}

\begin{figure}[t!]
\begin{center}
\mbox{(a)
\includegraphics[height=.23\textheight, angle =0]{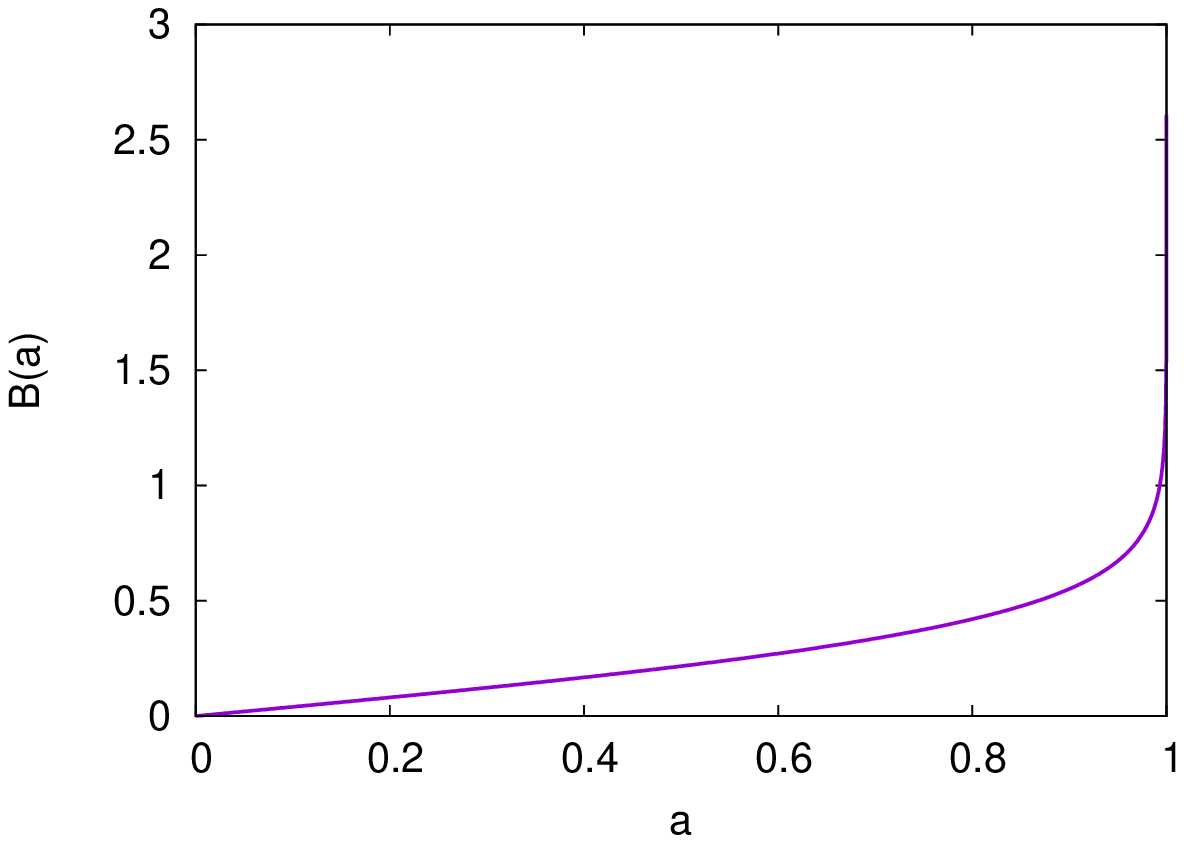}
(b)
\includegraphics[height=.23\textheight, angle =0]{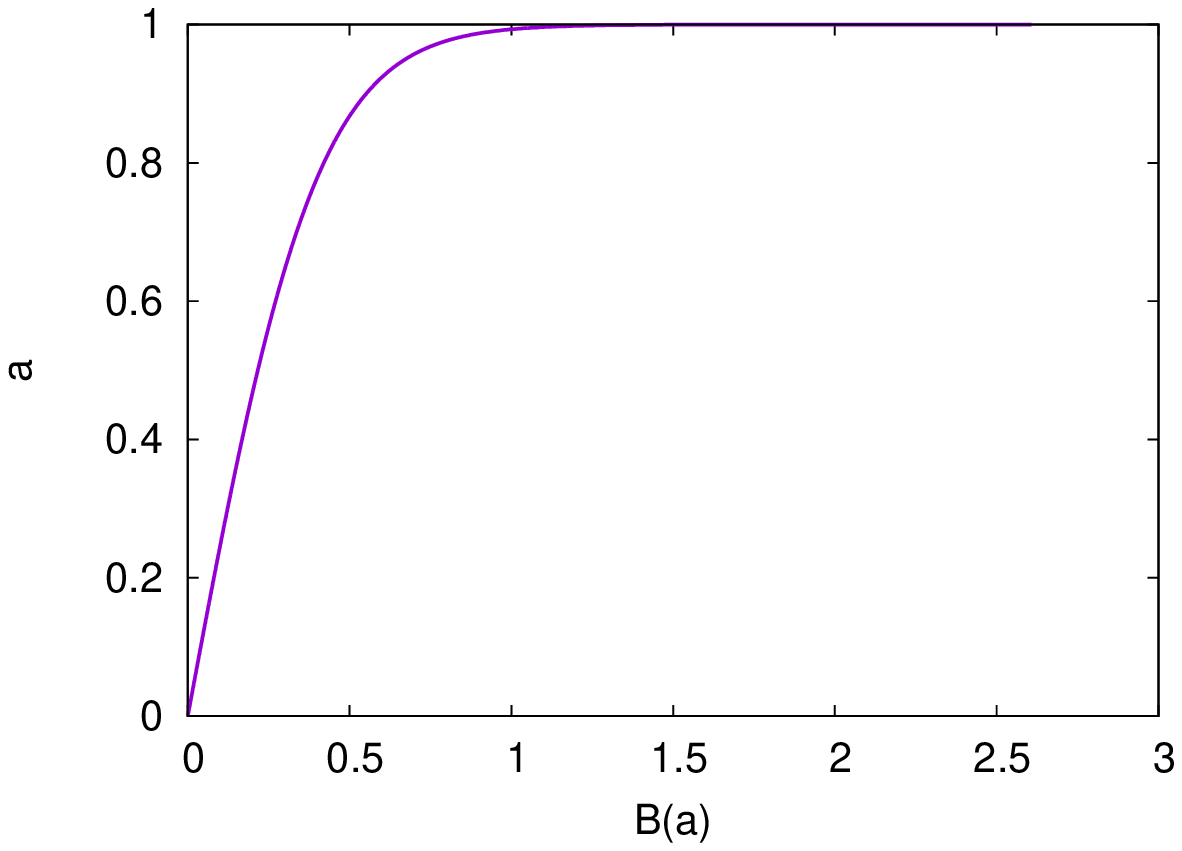}
}
\end{center}
\vspace{-0.5cm}
\caption{
Wormholes in STT-1 for negative values of $\beta$:
The function $B(a)$ (a) and its inverse (b).
\label{Fig10}
}
\end{figure}

Here we derive an analytical formula for the critical value of the scalar phantom charge
$\hat{Q}_\psi^{\rm cr}$. We start from Eq.~(\ref{eqQpsicrx1}).
In order to compute its rhs we note that $0< a^2 \le 1$ and 
consequently $0< a^2 e^{-\hat{\varphi}^2} \le 1$.
Excluding the equality, we can expand the square root,
\begin{equation}
\left\{1-a^2 e^{-\hat{\varphi}^2}\right\}^{-\frac{1}{2}}
= \sum_{n=0}^{\infty} c_n a^{2n} e^{- n \hat{\varphi}^2} \ ,  \ \ \ 
{\rm with} \ \ 
c_n= \frac{(2n-1)(2n-3)\cdots 1}{n!\, 2^n} \ .
\label{expsqr}
\end{equation}
Substitution in Eq.~(\ref{eqQpsicrx1}) leads to
\begin{equation}
\hat{Q}_\psi^{\rm cr}   =  
\pm\sum_{n=0}^{\infty} c_n a^{2n} \frac{1}{\pi}\int_{0}^{\infty}
e^{-(n+1/2)\hat{\varphi}^2} d\hat{\varphi} 
 =  \pm\frac{1}{2\sqrt{\pi}}\sum_{n=0}^{\infty} \frac{c_n}{\sqrt{n+1/2}} a^{2n} \ 
\label{eqQpsicrx1a}
\end{equation}
when evaluating the Gauss integral.
Multiplication of both sides with $a=\hat{D}/\hat{Q}_\psi^{\rm cr}$, 
with $\hat{D}=\sqrt{-\beta}D/\eta_0$
yields
\begin{equation}
\hat{D} =
\pm\frac{1}{2\sqrt{\pi}}\sum_{n=0}^{\infty} \frac{c_n}{\sqrt{n+1/2}} a^{2n+1}
 =: \pm B(a) =B(\pm a) \ . 
\label{eqQpsimcr2}
\end{equation}
Denoting by $B^{-1}$ the inverse function of $B$, 
we find
\begin{equation}
\pm \frac{\hat{D}}{\hat{Q}_\psi^{\rm cr}}
= B^{-1}\left(\hat{D}\right)
\Longleftrightarrow
\hat{Q}_\psi^{\rm cr} =\pm \frac{\hat{D}}{B^{-1}\left(\hat{D}\right)} \ .
\label{eqQpsimcr3}
\end{equation}
Thus, once the function $B(a)$ and its inverse are computed
(see Fig.~\ref{Fig10}),
Eq.~(\ref{eqQpsimcr3}) 
gives the upper and the lower bound for the scalar 
charge $\hat{Q}_\psi$ for given $\hat{D}(M,J)$ 
(and $\eta_0$, $\beta$).

A more convenient form of Eq.~(\ref{eqQpsimcr3}) is
\begin{equation}
\hat{Q}_\psi^{\rm cr}
= \hat{D}/B^{-1}\left(\hat{D}\right)
=:G\left(\hat{D}\right) \ .
\label{eqQpsimcr4}
\end{equation}

\end{document}